\documentclass[twocolumn]{aastex61}
\usepackage{amsmath}
\usepackage{lipsum}
\usepackage{graphics}
\usepackage[outdir=./]{epstopdf}

\newcommand\aastex{AAS\TeX}

\shorttitle{\aastex\ sample article}

\begin{document}

\title{AzTEC Survey of the Central Molecular Zone: Increasing Spectral Index of Dust with Density}


\author{Yuping Tang, Q. Daniel Wang, Grant W. Wilson}
\affil{Department of Astronomy, University of Massachusetts Amherst, Amherst, 01002, USA}


\begin{abstract}
The Central Molecular Zone (CMZ) of our Galaxy hosts an extreme environment analogous to that found in typical starburst galaxies in the distant universe. In order to understand dust properties in environments like our CMZ, we present results from a joint SED analysis of our AzTEC/Large Millimeter Telescope survey, together with existing \textit{Herschel} far-IR data on the CMZ, from a wavelength range of $160$ $\mu m$ to $1.1$ $mm$. We include global foreground and background contributions in a novel Bayesian modeling that incorporates the Point Spread Functions (PSFs) of the different maps, which enables the full utilization of our high resolution ($10.5''$) map at 1.1 $mm$ and reveals unprecedentedly detailed information on the spatial distribution of dusty gas across the CMZ. There is a remarkable trend of increasing dust spectral index $\beta$, from $2.0-2.4$, toward dense peaks in the CMZ, indicating a deficiency of large grains or a fundamental change in dust optical properties. This environmental dependence of $\beta$ could have a significant impact on the determination of dust temperature in other studies. Depending on how the optical properties of dust deviate from the conventional model, dust temperatures could be underestimated by $10-50\%$ in particularly dense regions. 
\end{abstract}


\section{Introduction} \label{sec:intro}

Observations have revealed that the central $\sim 200$ parsec region, or the main body of the so-called Central Molecular Zone (CMZ) of our Galaxy has an extreme gaseous environment, which may be common to the nuclear regions of many galaxies \citep{morris96}. The CMZ is characterized by dense ($n_{H_2} \gtrsim 10^{4}$ cm$^{-3}$), warm ($T \approx 60-100$ $K$) \citep{paglione98, oka07, ginsburg16} molecular gas with violent turbulent motions \citep{bally87, kauffmann17}. The magnetic fields \citep{morris15,pillai15} and the flux density of cosmic rays \citep{indriolo14,oka19} in the CMZ are larger here than anywhere else in the the Galactic disk. As a result, the CMZ hosts an environment with conditions similar to those observed in high redshift starburst galaxies \citep{kruijssen13, mills17}. 

The tight connections between gas conditions in the CMZ and high-redshift starburst galaxies highlight it as a template for verification/calibration of dust models in extreme environments. In modeling of high-redshift starburst galaxies \citep{blain02, casey14, popping17}, optical properties of dust grains are conventionally adopted from those inferred in the local environments. The dust absorption curve is normally simplified as a single power-law from far infrared (FIR) to submillimeter wavelengths, characterized by a spectral index $\beta=1.5-2$. Our ability to constrain dust properties in distant starburst galaxies is limited by the lack of spatial resolution \citep{casey12, magnelli12}. As a matter of fact, even in the local universe, studies of the spectral energy distributions (SEDs) of dust emission 
have not clearly established how the optical properties of dust vary in different environments. On small scales, $\beta \approx 1$ is observed in proto-planetary and proto-stellar disks \citep{draine06, kwon09} and is commonly attributed to $\gtrsim 1$ $mm$ size large grains. In dense molecular clouds and the diffuse ISM, a wide variety of $\beta$ is observed, from 0.8 to $>2$ \citep{dupac03, paradis11, juvela15}. The origin of this diversity is debated.
So far, observations suggest an anti-correlation between dust temperature and $\beta$ or a positive correlation between gas density $n_{H_2}$ and $\beta$ at long wavelengths $\lambda \gtrsim 500$ $\mu m$,  over the range from the diffuse ISM to cold dense clumps \citep{chen16,odegard16}. At short wavelengths $\lesssim 200-500$ $\mu m$, however, an inverse trend is observed \citep{ysard12},  i.e., a flattening of the dust absorption curve toward dense regions. 
It has been suggested that radiative transfer effects \citep{shetty09} and parameter degeneracies \citep{juvela13} could be responsible. The wavelength dependent change of $\beta$ is intriguing, and cannot be reproduced by classic models of dust growth \citep{ossenkopf94, kohler12, ysard12, ysard13}, which predict a negative $n_{H2}-\beta$ correlation extending to millimeter wavelengths. Recently, two new models have been proposed to solve this problem: a) accretion of small hydrogenated amorphous carbon onto large grains with updated optical properties of the hydrogenated amorphous carbons \citep{jones13, kohler15} and b) an intrinsic dependency of the dust absorption curve on the dust temperature \citep{meny07, paradis14}. These two models could be potentially distinguishable from observations, as the first scenario suggests a density dependency of $\beta$, and the second scenario suggests a temperature dependency. Nevertheless, observations with wide coverage in the $n_{H_2}-T_{dust}$ plane are required. It is also possible that turbulence \citep{hirashita09} is a factor affecting dust properties, which could potentially enhance shattering of large dust grains or suppress coagulation of small grains, especially in systems like the CMZ and high-redshift starburst galaxies where gas motions are extreme.  

Submillimeter/millimeter observations sampling the Rayleigh-Jeans tail of the dust SED are crucial for constraining the dust absorption curve. During Early Science Cycle 2 (ES2) for the Large Millimeter Telescope (LMT), we carried out a 20 hour survey of the dust continuum at 1.1 mm on the central $\approx 200$ pc of our Galaxy with the AzTEC bolometer array camera \citep{wilson08}. The AzTEC survey outperforms pre-existing FIR/submillimeter surveys (SPIRE/\textit{Herschel}, Bolocam/CSO, HFI/\textit{Planck}) 
with regard to spatial resolution (HPBW$=10.5''$). Existing studies of the dust emission in the CMZ are mostly based on the \textit{Herschel} Hi-GAL survey. \textit{Herschel}/SPIRE ($160-500$ $\mu m$) has comparatively low spatial resolution (HPBW$_{500\mu m}=36''$) and insufficient spectral coverage of the Rayleigh-Jeans tail of the dust SED. Adding a high-resolution survey at $1.1$ $mm$ to the current data set significantly enhances our capability to uncover small scale structures and place tighter limits on beta \citep{heyer18}. 

In this paper, we present the results of the AzTEC survey of the properties of dust in the CMZ, and study how parameters inferred from dust SEDs rely on different assumptions and priors. The structure of this paper is organized as follows. The observation strategy and data reduction for our AzTEC survey are briefly described in Section~\ref{sec:obs}, also described in Section~\ref{sec:obs} is how \textit{Herschel}-SPIRE/\textit{Planck}-HFI/CSO-Bolocam observations are processed and included into our study. A much more detailed discussion of both parts is presented in a separate paper on the data reduction of the AzTEC survey and the Bayesian analysis methodology of the multi-wavelength data (Tang et al.2020, hereafter Paper I). We extend the analysis to include the separation of the global background from the emission of the Galactic disk. In Section~\ref{sec:dust} we describe our detailed SED analysis. The results are presented in Section~\ref{sec:results_stmb}. The scientific implications are discussed in Section~\ref{sec:discussion}. In Section~\ref{sec:conclusion} we draw our conclusions.

\section{Observation \& Data Reduction}\label{sec:obs}

\subsection{AzTEC 1.1 mm Survey of the CMZ}

The AzTEC survey was conducted during ES2 for the 32 meter LMT, from Apr 17 to June 18, 2014, with a total integration time of $\approx 20$ hours. The survey covers the Galactic Center Region $l=[-0.7, 0.9]$, $b=[-0.6, 0.5]$, which roughly extends from Sgr B2 to Sgr C. The target field was mosaiced by square tiles, each observed with a raster-scan mode. The region immediately around Sgr B2 is excluded from this analysis due to Sgr B2's brightness, which exceeds 10-15 Jy and is subsequently not treated properly by our analysis due to the extreme dynamic range of flux and the possibility that the detector response is non-linear for the source. The optimal spatial resolution is $8.5''$ for the 32 meter LMT. However, since we adopted a high scanning speed of $200''/s$, we eventually obtain a beam size of $10.5''$ as a result of under-sampling. The $1\sigma$ noise level is about 15mJy/beam after a 20 hours integration time.

The raw data were reduced using the standard AzTEC analysis pipeline \citep{scott08}. 
We use iterative Principle Component Analysis (PCA) to remove correlated signals among bolometers, which are primarily contributed by the atmosphere, emissions from the telescope itself and non-Gaussian noises associated with the secondary mirror and back-end instruments.
PCA Cleaning is performed iteratively, until a conversion is reached such that the rms in a final noise map is consistent with no significant astronomical signal.

\begin{figure*}
\includegraphics[scale=0.5]{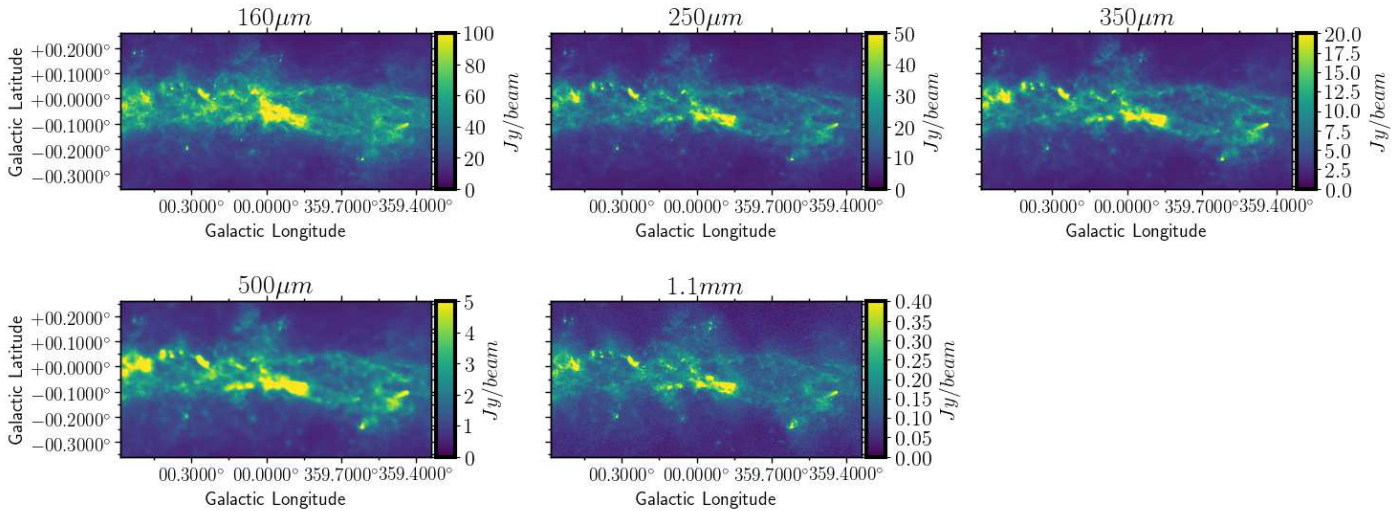}

\caption{160 $\mu m$-1.1 mm maps of the CMZ. The 1.1 mm map is combined from the AzTEC/LMT map, the Bolocam/CSO and the HFI/\textit{Planck} map.}
\label{fig:raw_map}
\end{figure*}


\subsection{Processing of \textit{Herschel}, \textit{Planck} and CSO/Bolocam maps}

We create a combined 1.1 mm map from the AzTEC 1.1 mm map, the \textit{Planck}/HFI $353$ GHz map (\textit{Planck} 2013 data release (PR1)) and the CSO/Bolocam 1.1 mm map \citep{aguirre11, ginsburg13}, to compensate for the large scale emission filtered out by the PCA cleaning in the AzTEC map. The \textit{Planck}/HFI ($353$ GHz or 850 $\mu m$) map is scaled to 1.1 mm to match the wavelength of the AzTEC \& Bolocam maps before merging. We apply pixel-by-pixel colors corrections and scaling factors from 850 $\mu m$ to 1.1 mm to the \textit{Planck} map inferred from best-fit $160$ $\mu m$-$850$ $\mu m$ SEDs. CO J=3-2 contamination has been removed from the \textit{Planck} map, using estimates provided by the \textit{Planck} Legacy Archive. The combined map is created using an approach described by \cite{faridani18}. This approach is mathematically equivalent to ”feathering”, an approach widely used for combining interferometer observations with single-dish observations, but is performed here in the “map domain” instead of the Fourier domain.

The $1\sigma$ statistical noise is $15$ mJy/beam in the AzTEC map, comparable to that in the Bolocam map. The noise in the \textit{Planck}/HFI map is negligible. We further apply a $10\%$ relative calibration uncertainty to the final 1.1 mm compound map, which accounts for beam variantions in the AzTEC maps. 

To construct dust SEDs, we further take advantage of existing \textit{Herschel} PACS/SPIRE $160$ $\mu m$, $250$ $\mu m$, $350$ $\mu m$ and $500$ $\mu m$ maps from the Hi-GAL survey \citep{molinari10}, which have been color-corrected using the Photometer Calibration Products from the ESA \textit{Herschel} Science Archive. The errors in the \textit{Herschel} maps are dominated by calibration uncertainties, which could be divided into relative calibration uncertainties and absolute calibration uncertainties. We adopt a relative calibration uncertainty of $2\%$ for all SPIRE bands, and a relative uncertainty of $5\%$ for the PACS $160$ $\mu m$ band \citep{bendo13, balog14}. We notice that some authors adopted more conservative estimates of the relative uncertainties for extended sources, inferred from comparisons between observations taken by \textit{Herschel}/PACS and those made by other facilities. (e.g. Spitzer/MIPS, AKARI, \citep{juvela15}). However, the fluctuations in the low surface brightness region of the PACS $160\mu m$ maps indicate that the relative uncertainty should be $<5\%$. Furthermore, it is dangerous to model absolute calibration offsets without an accurate knowledge of the dust absorption curve since these two ingredients are degenerate. Therefore, we ignore the absolute calibration uncertainties.

\subsection{Fore/Background Estimates} \label{sec:bk}

\begin{figure}
\includegraphics[scale=0.4]{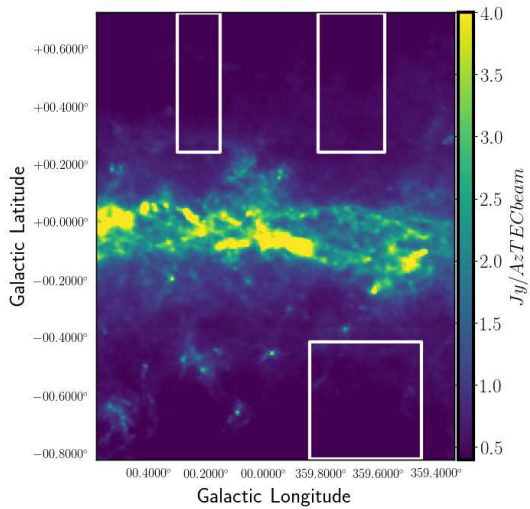}
\caption{\textit{Herschel} 500$\mu m$ map. The white rectangles show high-latitudes regions used to constrain fore/backgrounds.}
\label{fig:bkg}
\end{figure}

\begin{figure*}
\includegraphics[scale=0.35]{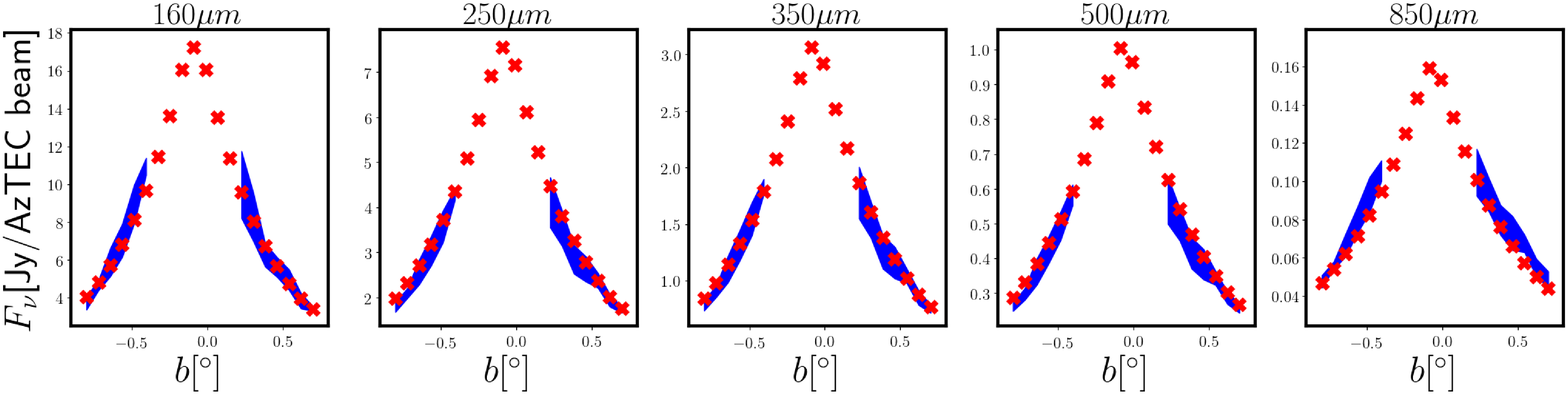}
\caption{The best-fit exponential background (red-crosses) compared with observed $\pm 1\sigma$ flux densities around the median values at different latitudes (blue-shaded area), data points are gathered 
from the 3 selected ``pure fore/background'' regions used for fitting.}
\label{fig:bkqual}
\end{figure*}

In order to separate the column densities of the CMZ clouds, $N_{cmz}$, from the foreground and background column densities $N_{fb}$, we assume that along each line of sight, the dust emission from the fore/background is only a function of Galactic latitude and that, for each, the column density and temperature exponentially decrease away from the Galactic plane. 

\begin{equation}\label{eq:pbk1}
N_{fb} = N_{fb,0} \times exp(- \frac{|b-b_{0N}|}{\sigma_{N}} )
\end{equation}

\begin{equation}\label{eq:pbk2}
T_{fb} = T_{fb,0} \times exp(- \frac{|b-b_{0T}|}{\sigma_{T}} )
\end{equation}

\noindent where $N_{fb,0}$ and $T_{fb,0}$ are peak column density and peak temperature, 
$b_{0\{N,T\}}$ and $\sigma_{\{N,T\}}$ are offsets and
scale heights, respectively. The spectral index $\beta$ is fixed to 1.8 for fore/background dust emission, which is derived from SED-fitting to the \textit{Herschel} $160-500$ $\mu m$ and \textit{Planck} 353 GHz maps degraded to the lowest resolution of the \textit{Planck} map.

The above model is constrained from three ``pure'' fore/background regions at high Galactic latitudes, which are shown in Figure~\ref{fig:bkg}. These low-flux regions are visually selected from the 500 $\mu m$ and 1.1 mm maps. $N_{fb,0}$, $T_{fb,0}$, $b_{0\{N,T\}}$ and $\sigma_{\{N,T\}}$ are derived by fitting $160-850$ $\mu m$ dust SEDs from these three regions, assuming that there is no CMZ component, pixel by pixel, after degrading every map to the lowest resolution at 850 $\mu m$. Figure~\ref{fig:bkqual} shows a comparison between the fitted model and the observed flux densities in the three defined ``pure'' fore/background regions. The blue shaded area shows $\pm 1\sigma$ flucations of the observed intensities at different latitudes. The best-fit model has $(N_{fb,0}[cm^{-2}]=10^{22.08}, b_{0N}=-0.05^{\circ}, \sigma_N=0.66^{\circ})$ and $(T_{fb,0}=24.8K, b_{0T}=-0.17^{\circ},\sigma_T=5.75^{\circ})$, notice that $T_{fb}$ is almost constant across the region. $\sigma_N=0.66^{\circ}$ corresponds to a scale height of $97$ pc, consistent with previous findings \citep{jones11, li18}.

This approach of the background subtraction is performed here in the $\{N,T\}$ space rather than on each flux map, such as that used by \cite{battersby11}. In this way, we take advantage of the knowledge that flux densities in different bands are correlated to follow an approximated modified black-body SED. 

\section{Modeling Dust Properties with Bayesian Analysis} \label{sec:dust}

In this section, we carry out a Bayesian analysis of the dust SEDs from the CMZ to explore the optical properties of dust grains. We adopt a forward modeling strategy to fit a dust model to multi-band maps, each diluted by a different instrumental PSF. The performance of this model-based deconvolution technique has been demonstrated in Paper I with a model of single temperature modified black-body (hereafter STMB). In this work, we extend and optimize this analysis to improve the estimation of physical parameters intrinsic to the CMZ.

\subsection{Single Temperature Dust Model}
Here we briefly recap the STMB model that is detailed in Paper I. This relies on three parameters: dust temperature $T$,  column density grid $N_{H_2}$ and dust spectral index $\beta$. The surface brightness $F_{i}(\nu_j)$ at pixel(i) and frequency $\nu_j$ is given by:

\begin{equation} \label{eq:modbk}
F_{i}(\nu_j) = [1-exp(-\tau_{i,\nu_j})] B_{\nu_j}(T_{i}) \Omega_j
\end{equation}

\noindent where $\Omega_j$ is the beam area in the jth band. 
$B_{\nu_j}(T_{i})$ is the Planck function. ${\tau}_{i,\nu_j}$ is the optical depth at frequency $\nu_j$, which is given by:

\begin{equation} \label{eq:tau}
\tau(i,\nu_j) = \kappa_0 (\frac{\nu_j}{\nu_0})^{\beta_{i}} \mu m_{H} \times {N_{H_2}}_{i} \times 1\%   
\end{equation}

\noindent where $\kappa_0$ is the absorption cross section per unit mass at frequency $\nu_0$. We adopt $\kappa_0=1.37$ cm$^2/g$ and $\nu_0=c/1000$ $\mu m$ from \cite{ossenkopf94} for coagulated dust grains with thin ice mantles (their Table 1). We also adopt a mean molecular weight $\mu=2.8$ per H$_2$ molecule from \cite{kauffmann08} and a dust-to-gas mass ratio of $1\%$ to convert from $N_{H_2}$ to column dust mass density. This model is not restricted to an optically thin approximation ($\kappa \propto \nu^{\beta}$).

The raw flux map $\mathbf{F(\nu_j)}$ calculated above is diluted to the instrumental resolution of each wavelength band to match the data:
\begin{equation}
\mathbf{Model(\nu_j)} = \mathbf{F(\nu_j)} \otimes \mathbf{beam_{j}}
\end{equation}

\noindent where $\otimes$ refers to convolution. All beams profiles are approximated as Gaussian profiles. The Full Width Half Maximum (FWHM) of the beams are $13.6''$ at 160 $\mu m$, $23.4''$ at 250 $\mu m$, $30.3''$ at 350 $\mu m$, $42.5''$ at 500 $\mu m$ and $10.5''$ at 1.1 mm, respectively. The beam sizes of the PACS/SPIRE maps are larger than their nominal values \citep{traficante11}, which is due to the high scanning speed adopted by the Hi-Gal survey.

\subsection{STMB with Fore/Background Subtraction}

The total flux along any line of sight is the sum of the CMZ and its fore/background. With the self-absorption being accounted for, the total flux is:

\begin{equation}\label{eq:Ntot1}
F_{tot} = F_{bg} \times exp(-\tau_{cmz}-\tau_{fg}) + F_{cmz} \times exp(-\tau_{fg}) + F_{fg} 
\end{equation}

\noindent where $F_{bg}$/$F_{cmz}$/$F_{fg}$ are the intrinsic intensities of the background/CMZ/ foreground components, respectively. Because our concerned Galatic latitude range is small, we assume that the foreground and background intensities along each line of sight are identical: $F_{bg} = F_{fg} = (1-exp(-\frac{1}{2}\tau_{fb})) B_{\nu}(T_{fb})$, $\tau_{bg}=\tau_{fg}=\frac{1}{2}\tau_{fb}$.

For MCMC sampling, it is more convenient to use the integrated column density, $N_{tot}$ along each line of sight as a free parameter, instead of $N_{cmz}$. Then we have:

\begin{equation}\label{eq:ftot}
     \resizebox{0.5\textwidth}{!}{$
     F_{tot}
     = \begin{cases}
     F(N_{tot}, \{T,\beta\}_{fb}), \text{if $N_{tot} <= N_{fb}$ $\boldsymbol{(a)}$}, \\
     \\
     F(\{T,N,\beta\}_{bg}) \times exp(-\tau_{cmz}-\tau_{fg}) \\ 
     +  F(\{T,N,\beta\}_{cmz}) \times exp(-\tau_{fg}) \\
     + F(\{T,N,\beta\}_{fg}), \text{if $N_{tot} > N_{fb}$ $\boldsymbol{(b)}$}, \\
     \end{cases}
     $}
\end{equation}

\noindent where $F(\{T,N,\beta\}) = [1-\tau(N,\beta)] B_{\nu_i}(T)$ while $\tau$ is the total internal opacity. $N_{cmz} = N_{tot} - N_{fb}$, $\tau_{bg} = \tau_{fg} = \frac{1}{2}\tau_{fb}$, $N_{bg} = N_{fg} = \frac{1}{2}N_{fb}$, $T_{bg} = T_{fg} = T_{fb}$ and $\beta_{bg} = \beta_{fg} = 1.8$. Note that fore/background fluxes are not completely fixed to the intensities calculated from Eq~\ref{eq:pbk1} and Eq~\ref{eq:pbk2}. In Eq~\ref{eq:ftot} $\boldsymbol{(a)}$, while $T$ is always fixed to the values $T_{fb}$ inferred from the best-fit Eq~\ref{eq:pbk2} and $\beta$ is to $1.8$, $N_{tot}$ could take values smaller than $N_{fb}$. In other words, low column density cells are not elevated to the values inferred from Eq~\ref{eq:pbk1}.

\subsection{MCMC Analysis: Sampling Strategy} \label{MCMCpost}

In Paper I, we have demonstrated a Markov chain Monte Carlo (MCMC) approach for a model-based deconvolution of multi-band maps. A key ingredient of this procedure is a smoothness prior. In this section, we start by recapping the smoothness prior that is defined in Paper I, and then propose a more generalized form of smoothness prior that is applicable to the STMB model with the fore/background subtraction. 

As is well known, deconvolution normally results in over-fitting to noise, which manifests as high-frequency fluctuations among neighboring cells in the best-fit maps. We employ regularized Bayesian inference~\citep{warren03} to mitigate this issue. Briefly, we adopt a simple form of smoothness prior based on the local gradients of the parameter $x_{ix,iy}$ to be sampled:

\begin{equation} 
ln(P(x_{ix,iy})) = ln(L(x_{ix,iy})) + P_1
\end{equation}

\noindent in which:

\begin{equation} \label{eq:smprior}
P_1 =  -\frac{1}{2} \frac{\sum\limits_{j=-1, 1}(x_{ix+j,iy} - x_{ix,iy})^2 + \sum\limits_{k=-1, 1}(x_{ix,iy+k} - x_{ix,iy})^2}{{2\sigma_1^2}}
\end{equation}

\noindent Here, $ln(P(x_{ix,iy}))$ and $ln(L(x_{ix,iy}))$ are the logarithms of the full conditional posterior and likelihood for parameter $x_{ix,iy}$, where $\mathbf{x}$ is $\{N_{H_2},T,\beta \}$ while $P_1$ is an \textit{a priori} smoothness of $\mathbf{x}$ and is estimated with gradients among all adjacent cells in a grid. $\sigma_1$ is user-defined and could be viewed as an \textit{a priori} mean standard deviation of adjacent cell-cell differences. In Paper I, we explore the choices of $\sigma_1$ and show that, as demonstrated by spectral density analysis, for a broad range of $\sigma_1$, the performance of this regularized deconvolution approach is superior to that of the conventional approach that involves degrading every image to the lowest resolution. 

For an STMB with the fore/background subtraction, however, the above prior has two issues. First, since there is no reason to assume any continuity between a CMZ component and a fore/background component, $P_1$ for $T_{cmz}$ or $\beta_{cmz}$ should be defined only among pairs of cells both having a CMZ component (i.e., $N_{tot}>N_{fb}$). And since $P_1$ has a negative contribution to the posterior, it yields a bias against having a CMZ component in every cell. Furthermore, in low density regions, where a cell having a CMZ component can have less than 4 neighbors which also have a CMZ component, Eq~\ref{eq:smprior} has less constraining power. To solve both issues, we propose a more generalized form of smoothness prior, calculated from average gradients upon a $n \times n$ block centered at each cell to be sampled:    

\begin{equation} \label{eq:smprior_n}
P_n = -2 \frac{\sum\limits_{j=-\frac{n-1}{2}}^{\frac{n-1}{2}} \sum\limits_{k=-\frac{n-1}{2}}^{\frac{n-1}{2}} (x_{ix+j, iy+k} - x_{ix,iy})^2 w_{j,k}/(2\sigma_n^2)} {\sum\limits_{j=-\frac{n-1}{2}}^{\frac{n-1}{2}} \sum\limits_{k=-\frac{n-1}{2}}^{\frac{n-1}{2}} f_{cmz,j,k}} 
\end{equation}

\noindent where n is an odd number, while $w_{j,k}$ is defined as:

\begin{equation}
w_{j,k} = \begin{cases} \frac{1}{\sqrt{j^2+k^2}}, \text{if $N_{tot,0,0}>N_{fb,0,0}$ and $N_{tot,j,k}>N_{fb,j,k}$}, \\
          0, \text{if $N_{tot,0,0}<=N_{fb,0,0}$ or $N_{tot,j,k}<=N_{fb,j,k}$}, \\ 
          \end{cases}
\end{equation}

\noindent and $f_{cmz}$ is:

\begin{equation}
f_{cmz,j,k} = \begin{cases} 1, \text{if $N_{tot,0,0}>N_{fb,0,0}$ and $N_{tot,j,k}>N_{fb,j,k}$}, \\
          0, \text{if $N_{tot,0,0}<=N_{fb,0,0}$ or $N_{tot,j,k}<=N_{fb,j,k}$}, \\ 
          \end{cases}
\end{equation}

\noindent Notice that from Eq~\ref{eq:smprior} to Eq~\ref{eq:smprior_n}, the scaling factor changes from $\frac{1}{2}$ to $2$. In Eq~\ref{eq:smprior}, the factor of $\frac{1}{2}$ accounts for the fact that each pair is counted twice when $P_1$ is summed over all cells. In Eq~\ref{eq:smprior_n}, the factor of $2$ ensures that $\sigma_n$ is defined in a comparable fashion to $\sigma_1$, since gradients in Eq~\ref{eq:smprior} are summed over all adjacent pairs, which have a total number of $\approx 2N_{cell}$.

Throughout this study, we use both $P_1$ and $P_n$ defined above to relieve over-fitting during forward modeling. $P_n$ is adopted only for the STMB model with the fore/background subtraction. The smoothness priors for different models discussed in this paper are summarized in Table~\ref{tab:prior}, where $\sigma_1$ corresponds to $P_1$ and $\sigma_n$ ($n>1$) corresponds to $P_n$.

\begin{table*}[]
\centering
\caption{Parameters and Smoothness Prior for Different Models}
\label{tab:prior}
\resizebox{\textwidth}{!}{
\begin{tabular}{lcr}
  Model & Free Parameters & Smoothness Prior  \\
  \hline
 STMB & $lg(N_{H_2})$, $ln(T)$, $\beta$  & $\sigma_{1,lg(N)}=0.1, \sigma_{1,\beta}=0.2^a$  \\
 STMB with fore/background subtracted & $lg(N_{tot})$, $ln(T_{cmz})$, $\beta_{cmz}$  & $\sigma_{1,lg(N_{tot})}=0.1$, $\sigma_{5, \beta_{cmz}}=0.1$   \\
 STMB with fore/background subtracted, $7''^{b}$ & $lg(N_{tot})$, $ln(T_{cmz})$, $\beta_{cmz}$  & $\sigma_{1,lg(N_{tot})}=0.05$, $\sigma_{5, ln(T_{cmz})}=0.05$, $\sigma_{5, \beta_{cmz}}=0.05$ \\ 
 STMB with a multivariate prior & $lg(N)$, $ln(T)$, $\beta$, $\mu$, $\Sigma$ & $\sigma_{1,lg(N)}=0.1, \sigma_{1,\beta}=0.2$ \\
 STMB with a broken power-law absorption curve & $lg(N)$, $ln(T)$, $\beta_2$, $\lambda_{bk}$ & $\sigma_{1,lg(N)}=0.1$, $\sigma_{1, \beta_1}=0.1$, $\sigma_{1, \lambda_{bk}[\mu m]}=20$ \\
 TLS model & $lg(N_{H_2})$, $ln(T)$ & $\sigma_{1,lg(N)}=0.1$, $\sigma_{1,ln(T)}=0.2$ \\
 \hline
\end{tabular}
}
\noindent (a) $\sigma_{1,x}$ indicates a prior defined by Eq~\ref{eq:smprior}, while $\sigma_{n,x}(n>1)$ a prior defined by Eq~\ref{eq:smprior_n}. \\
\noindent (b) The pixel/cell size is $14''$ unless specified.
\end{table*}

\subsection{Hierarchical Bayesian Model} \label{sec:hiebayes}

A common problem encountered in physical modeling is parameter degeneracy. The global distribution of the estimated parameters could be viewed as a convolution of their natural distribution with the probability distributions of their estimated values propagated from measurement uncertainties. Given that measurement uncertainty is always present, parameter degeneracy leads to correlated probability distributions, which dilute the apparent distribution of the best-fit parameters toward a false correlation. In SED analysis with a STMB model, $T$ and $\beta$ are known to have a high degeneracy, which manifests as a banana-shaped posterior distribution. This makes it difficult to recover the intrinsic $T-\beta$. \cite{juvela13} examined several existing techniques aiming to recover the intrinsic $(T,\beta)$ relation and concluded that all techniques suffer from some bias. 

Hierarchical Bayesian Analysis has been proposed to remedy the intrinsic correlation between $T$ and $\beta$, by implementing the natural distribution of parameters as a prior to the model \citep{kelly12, galliano18}.  Following \cite{kelly12}, we adopt a multivariate Student-t distribution as a prior for the $T-\beta$ distribution. The posterior can be written as:

\begin{equation} 
\resizebox{0.45\textwidth}{!}{$
P(\mu,\Sigma | D) = \prod_{i} P(D | \mathbf{x_i}) P(\mathbf{x_i} | \mu, \Sigma) \times P(\mu, \Sigma)$} 
\end{equation}

\begin{equation} \label{eq:student}
\resizebox{0.5\textwidth}{!}{$
P(\mathbf{x_i} | \mu, \Sigma) \propto \frac{1}{|\Sigma|^{1/2}} \times [1 + \frac{1}{d}(\mathbf{x_i} - \mu)^{T} \Sigma^{-1} (\mathbf{x_i} - \mu)]^{-(d+2)/2}
$}
\end{equation}

\begin{equation}
\mathbf{x_i}=(ln(T_i), \beta_{i})
\end{equation}

\noindent where D is the data, and $\mu$ is the global mean of $\mathbf{x_i}$. $\Sigma$ is the covariance matrix of $\mathbf{x_i}$. When d+1 samples are drawn from a normal distribution, Eq~\ref{eq:student} can be viewed as a distribution of the deviation of the sample mean from the true mean, divided by a normalized sample standard deviation. Here d is the degrees of freedom and is arbitrarily set to 8 following \cite{kelly12}. A Student-t distribution with smaller degrees of freedom has a larger portion of outliers relative to a normal distribution.

The covariance matrix $\Sigma$ can be decomposed as:

\begin{equation}
\Sigma = SRS
\end{equation}

\begin{figure}
\includegraphics[scale=0.5]{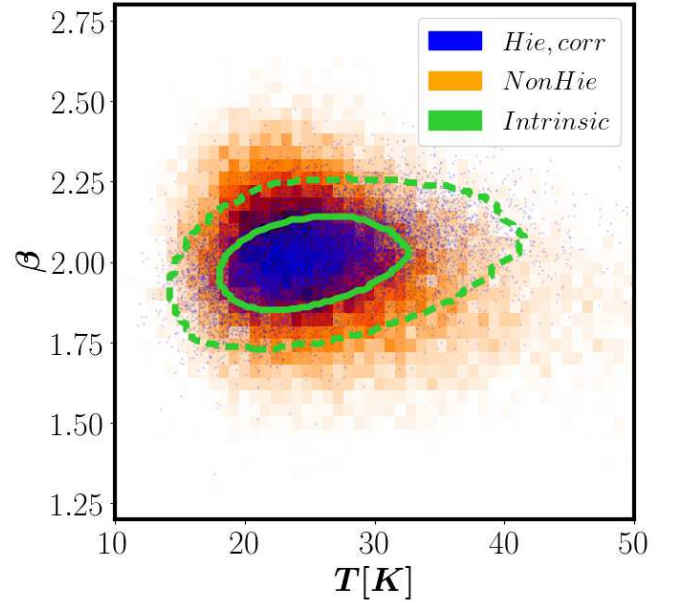}
\caption{Distribution of $T$ and $\beta$ recovered from hierarchical (blue) and non-hierarchical (2d histograms) Bayesian analysis. The SEDs are simulated from a 3x3 multivariate normal distribution for $(lg(N),ln(T), \beta)$, with $(\mu_{lg(N[cm^{-2}])}=22.5, \mu_{ln(T[K])}=Ln(20), \mu_{\beta}=2)$, $(\sigma_{lg(N[cm^{-2}])}=0.2, \sigma_{ln(T[K])}=0.2, \sigma_{\beta}=0.1)$ and $(\rho_{lg(N), Ln(T)}= -0.5, \rho_{lg(N), \beta}=0.5, \rho_{ln(T), \beta}=0.3$. The 2D histograms show the distribution estimated from a non-hierarchical Bayesian model, which show an apparent anti-correlation, the blue dots mark values estimated from a hierarchical Bayesian analysis, with $(ln(T), \beta)$ following a prior of a 2x2 multivariate Student-t distribution. The green contours correspond to the simulated marginal distribution of $(ln(T), \beta)$.}
\label{fig:hie_simu}
\end{figure}

\noindent where $S$ is the diagonal matrix of the standard deviations and R is the correlation matrix.

With $P(ln(T_{i}), \beta_{i} | \mu, \Sigma)$ as an additional prior, we have 5 hyperparameters: 
$\mu_{ln(T)}$, $\mu_{\beta}$, $\sigma_{ln(T)}$, $\sigma_{\beta}$ and $\rho_{ln(T), \beta}$, which are sampled along with 
$(lg(N_i), ln(T_i), \beta_i)$ using a slice-within-Gibbs strategy. Since the covariance matrix $\Sigma$ is a 2x2 matrix, 
it is always positive-definite as long as $-1< \rho_{ln(T), \beta}< 1$. We can simply place
a uniform prior on $\rho_{ln(T), \beta}$ between $-1$ and $1$. We further give uniform priors on the rest of parameters: 
$\mu_{ln(T[K])} \in (ln(5), ln(60))$, $\mu_{\beta} \in (0.5, 3.0)$, $\sigma_{ln(T[K])} \in (0.02, 0.4)$, and $\sigma_{\beta} \in (0.02, 0.4)$. 

An illustration of this hierarchical Bayesian analysis is shown in Figure~\ref{fig:hie_simu}. Here, we simulate a sample of $22400$ dust SEDs from a 3x3 multivariate normal distribution, with $(\mu_{lg(N[cm^{-2}])}=22.5, \mu_{ln(T[K])}=ln(20), \mu_{\beta}=2)$, $(\sigma_{lg(N[cm^{-2}])}=0.2, \sigma_{ln(T[K])}=0.2, \sigma_{\beta}=0.1)$\footnote{$lg$ means base-10 $log$} 
and $(\rho_{lg(N), ln(T)}= -0.5, \rho_{lg(N), \beta}=0.5, \rho_{ln(T), \beta}=0.3$. The signal-to-noise ratios are identical to our observed data. $ln(T)$ and $\beta$ are simulated to follow a positive correlation with $\rho_{ln(T), \beta}=0.3$. The best-fit distribution of $ln(T)$ and $\beta$ derived from a regular Bayesian analysis shows an apparent anti-correlation. For this simulation, we adopt a multivariate prior (2x2 multivariate normal distribution) that correctly characterizes the natural distribution (3x3 multivariate normal distribution) and is able to accurately recover the intrinsic correlation between $T$ and $\beta$. We caution that the natural $T-\beta$ distribution in molecular clouds might not follow a multivariate bell-shape distribution as we simulated. However, based on a magnetohydrodynamical (MHD) simulation of molecular clouds, \cite{juvela13} has shown that, multivariate prior could retain the information of the correlation coefficient under
reasonable noise levels.

\section{Results: Single Temperature Modified Blackbody Model (STMB)} \label{sec:results_stmb}

\begin{figure*}
\begin{center}
\includegraphics[scale=0.5]{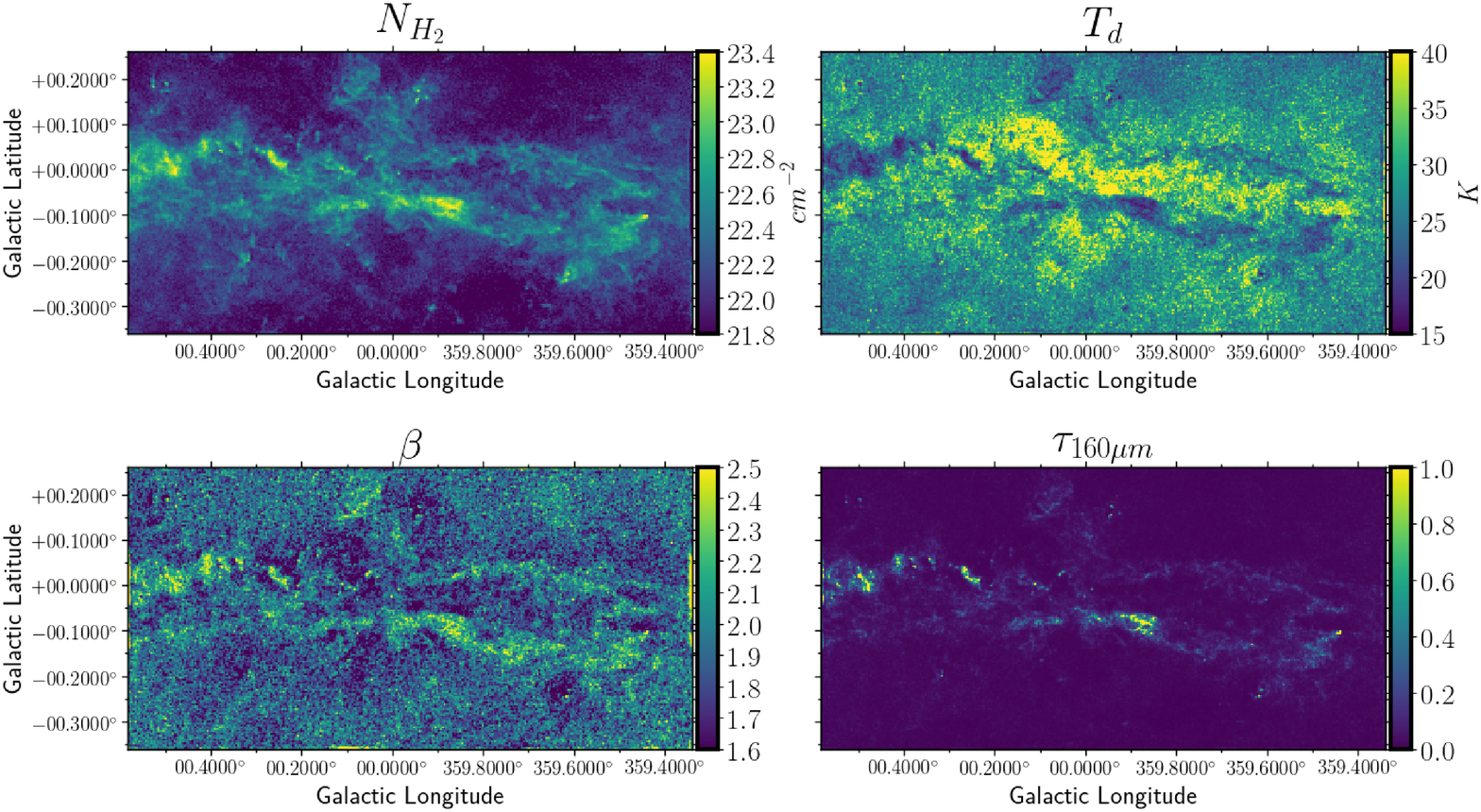}
\includegraphics[scale=0.5]{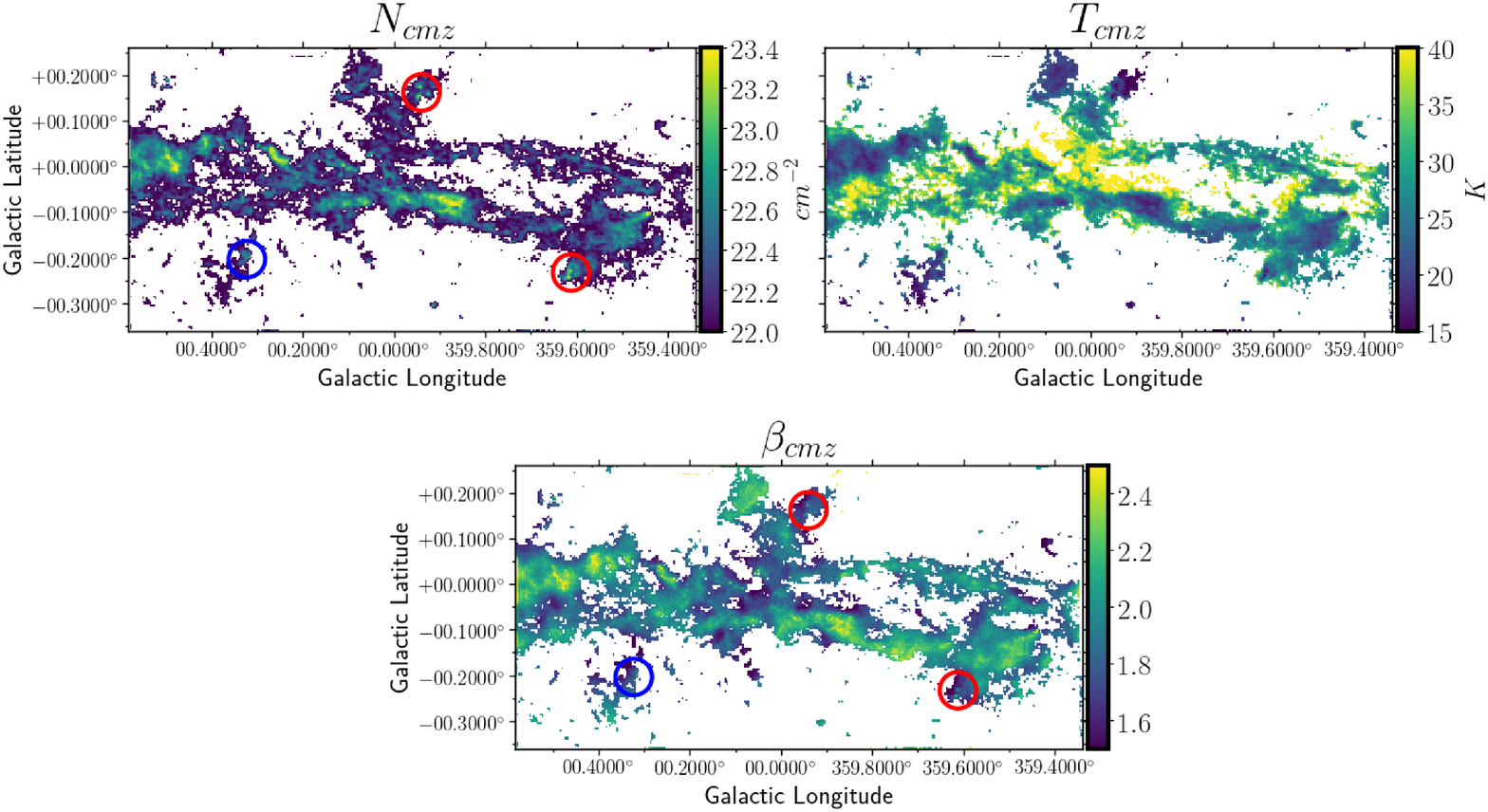}
\caption{Upper four panels: Best-fit temperatures (upper left), equivalent hydrogen column densities (upper right), $\beta$ (lower left) and
optical depths at $160$ $\mu m$ (lower right), obtained with a single temperature, modified black-body approximation with no fore/background subtraction. The pixel/cell size is $14''$. The last map shows the integrated optical depths along the line of sight at $160$ $\mu m$, which are inferred from $N_{H_2}$, $\kappa_0$ and $\beta$. The red circles enclose two regions identified to be dominated by foreground objects, as suggested by their $V_{lsr}$, while the blue circle marks an object possibly associated with the HII region SH-20. Lower three panels: The same as the first three maps in the upper panels, but after the fore/background subtraction}
\label{fig:maps_singT}
\end{center}
\end{figure*}

\begin{figure*}
\begin{center}
\includegraphics[scale=0.8]{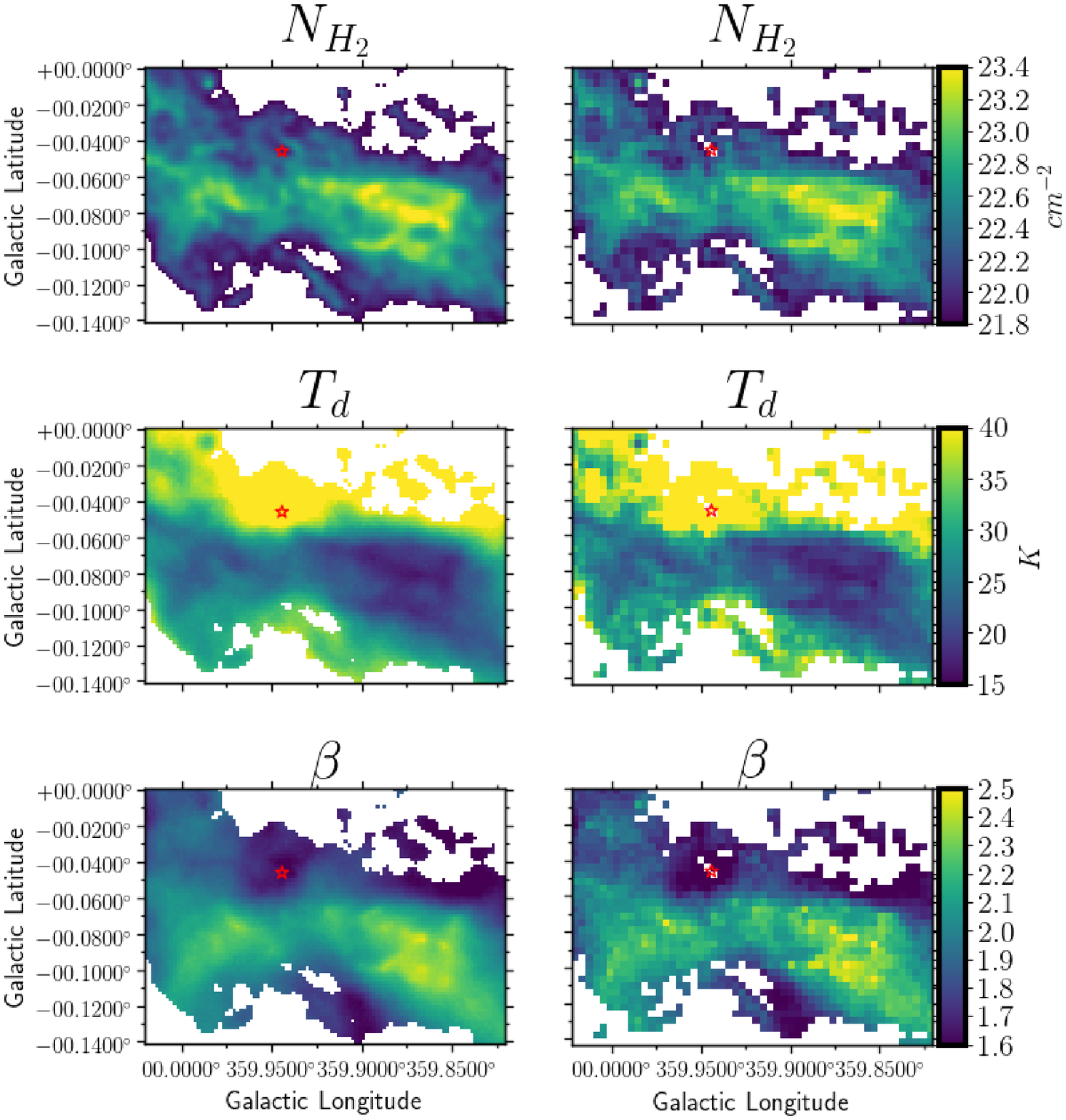}

\caption{A close-up comparison between best-fit maps with a 7'' pixel/cell size and those with a 14'' pixel/cell size. The red-star indicates Sgr A*, the data around which are largely contaminated by its nonthermal emission.}

\label{fig:7_14_cp}
\end{center}

\end{figure*}

The product maps of $T$, $N_{H_2}$ and $\beta$ before and after fore/background subtraction are shown in Figure~\ref{fig:maps_singT}.
Overall, the ranges of $T$ and $N_{H_2}$
are similar to those derived by \cite{molinari11} with DUSTEM. The temperatures of dense clouds are typically $\lesssim 20K$, 
and the peak column density $N_{H_2}$ is $\approx 10^{23.5}$ cm$^{-2}$. Below $N_{H_2} \approx 10^{22}$ cm$^{-2}$, 
the fluxes are dominated by fore/background emission. 

As discussed in Paper I, the effective resolution in each best-fit map derived by our model-based deconvolution approach is between $10.5''$ and $42.5''$ and is different from parameter to parameter. While $N_{H_2}$ strongly depends on the combined 1.1 mm map, which has the highest resolution, T and $\beta$ are more dependent on lower resolution maps. The final achieved resolution also relies on the choice of the smoothness prior. With a smaller cell size, the number of free parameters increases, but there is no extra information on sub-PSF scales. In other words, the effective degrees of freedom are reduced. Therefore, stronger smoothness priors are required to avoid overfitting. We have examined the best-fit maps after the fore/background subtraction with two different configurations, one configuration with a smaller cell/pixel size ($7''$) and strong priors ($\sigma_{1, lg(N)}=0.05$, $\sigma_{5,ln(T)}=0.05$, $\sigma_{5,\beta}=0.05$), a second one with a larger cell/pixel size ($14''$) and weak priors ($\sigma_{1, lg(N)}=0.1$, $\sigma_{5,\beta}=0.1$). A close-up comparison of the two results is shown in Figure~\ref{fig:7_14_cp}. On scales larger than $42.5''$, there is no apparent difference. Since this particular study focuses on the global distributions of dust properties in the CMZ, we determine to use a cell/pixel size of $14''$ for all models so as to avoid bias induced by the priors.

\begin{figure*}
\begin{center}
\includegraphics[scale=0.32]{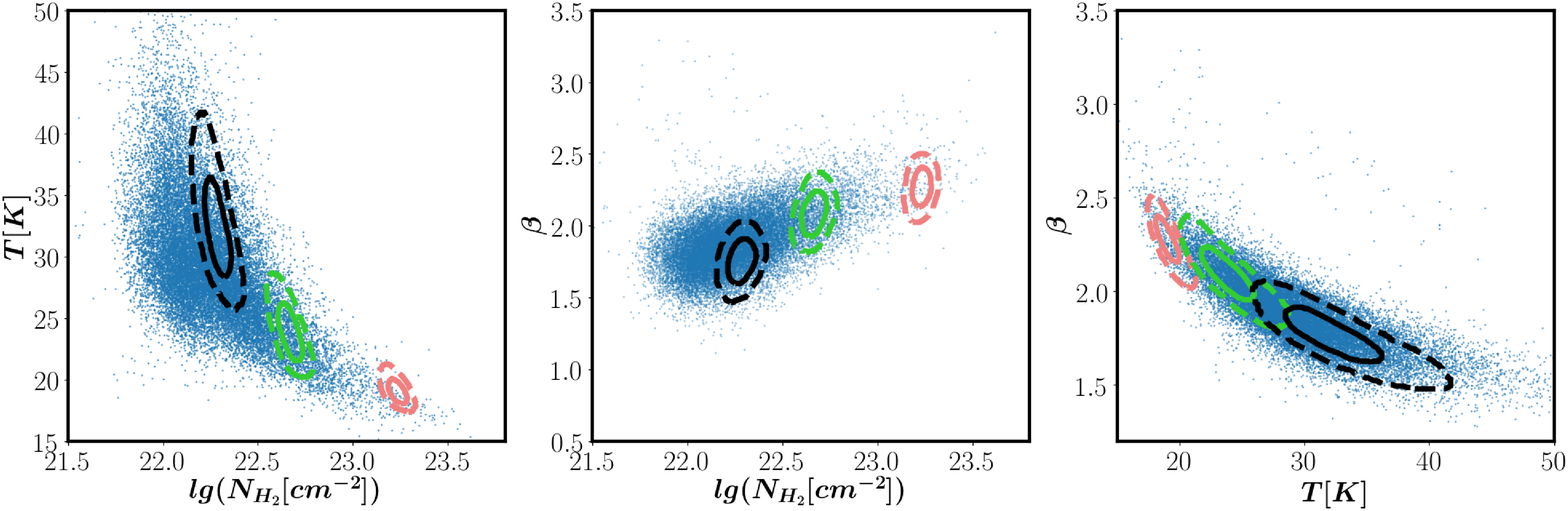}
\includegraphics[scale=0.32]{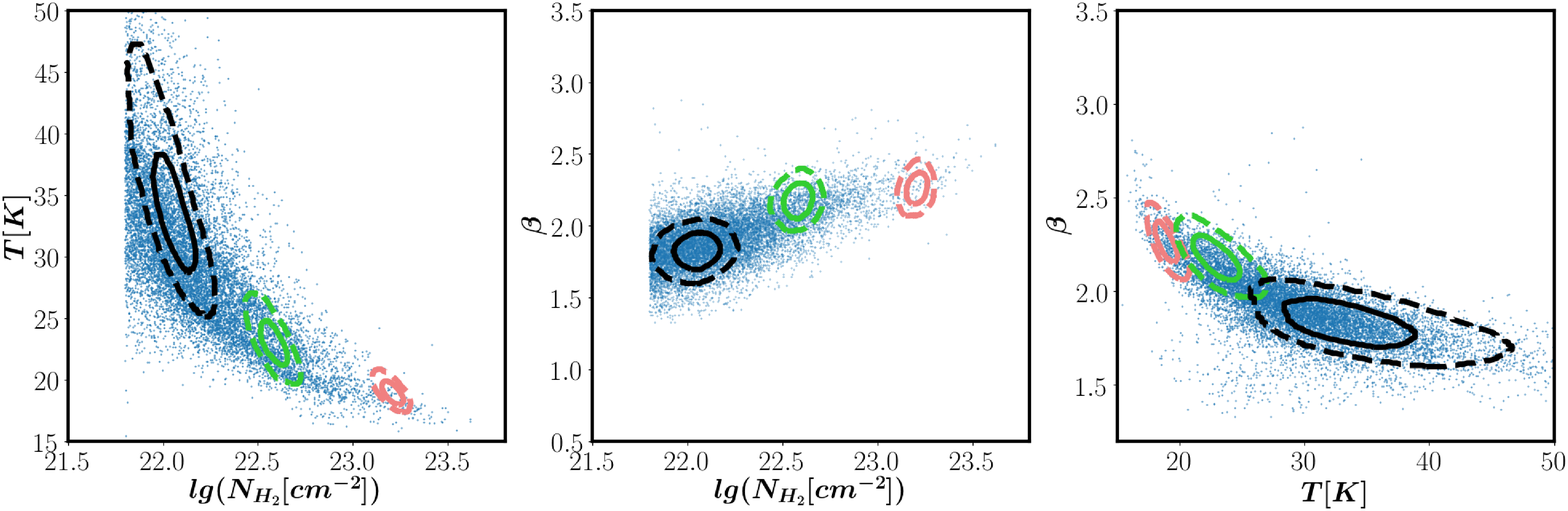}
\caption{Upper panels: Correlations between parameters derived from the best-fit STMB model. Only the low Galactic latitude ($-0.2^{\circ}<b<0.1^{\circ}$) cells (blue dots) are used in this derivation. Also plotted are the 1 and 3 $\sigma$ confidence contours of three typical sampled posteriors at different representative locations in the parameter space. Lower panels: Similar plots after the fore/background subtraction, using only cells with $N_{H_2, cmz}>10^{21.8}$.}

\label{fig:corr}
\end{center}
\end{figure*}

On large scales, there are two pronounced correlations: a negative correlation between $N_{H_2}$ and $T$ and a positive correlation between $N_{H_2}$ and $\beta$. 
The marginalized distributions of the best-fit $T$, $N_{H_2}$ and $\beta$ are plotted in Figure~\ref{fig:corr}. In each panel we also plot three typical projected sampled posteriors at different locations in the parameter space. Cells in high galactic latitudes ($b<-0.2^\circ$ or $b>0.1^\circ$) are excluded. Measurement uncertainties are partially responsible for the apparent correlation between estimated $T$ and $\beta$ , which propagates into a banana-shaped posterior distribution. However, the sampled posterior distributions suggest that a genuine anti-correlation between $T$ and $\beta$ is present. The hierarchical Bayesian analysis also supports an intrinsic $T$-$\beta$ anti-correlation. Figure~\ref{fig:hie_singT} shows the $T-\beta$ distribution derived by modeling the natural $T-\beta$ distribution as a multivariate Student-t prior distribution (Section~\ref{sec:hiebayes}). 
The $T-\beta$ distribution does not significantly change other than a reduction of high-temperature cells. 
The estimated correlation coefficient $\rho_{ln(T), \beta}=-0.74$ indicates a strong anti-correlation. On the other hand, through the same analysis, we find that $\rho_{lg(N_{H_2}), \beta}=0.89$, indicating an even stronger correlation between $N_{H_2}$ and $\beta$.

\begin{figure*}
\begin{center}
\includegraphics[scale=0.34]{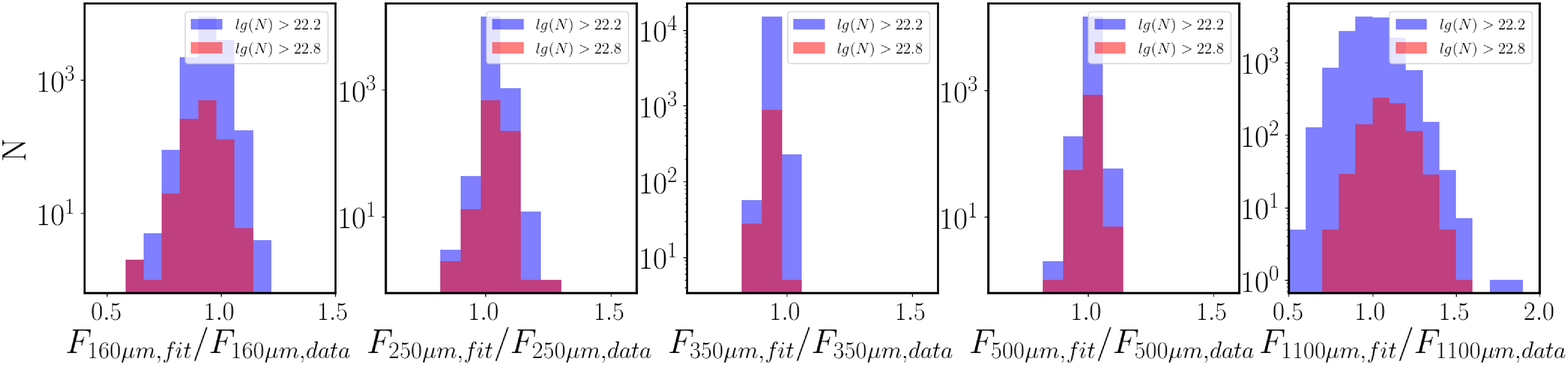}
\includegraphics[scale=0.34]{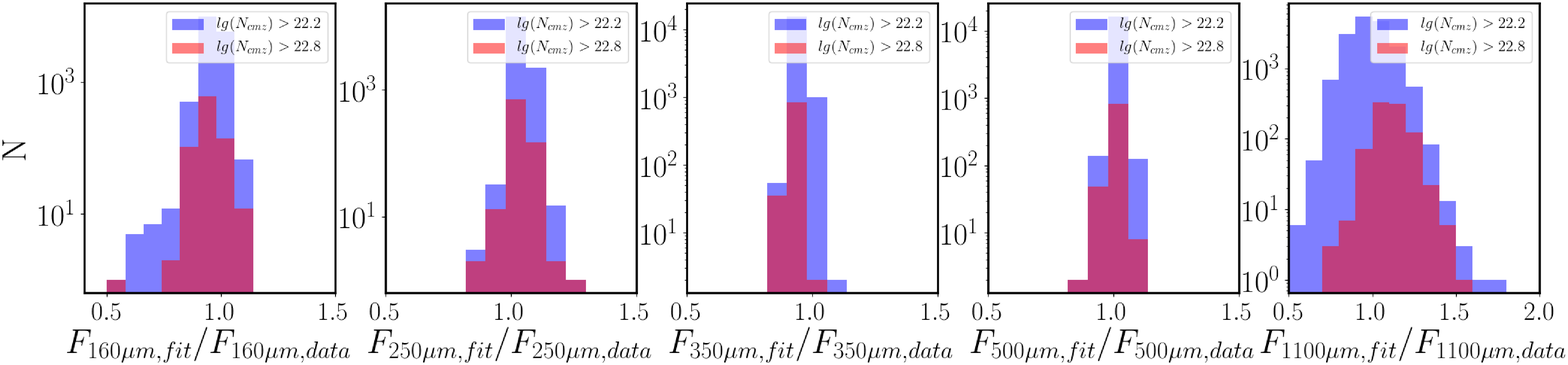}
\caption{Histograms of the ratios $F_{fitted}$/$F_{observed}$ without(upper) and with(lower) fore/background subtraction.
Blue bars correspond to all pixels satisfying ($lg(N_{H_2}[cm^{-2}])>22.2$) and red bars correspond to high column densities pixels only ($lg(N_{H_2}[cm^{-2}])>22.8$). 
For results without fore/background subtraction, pixels with high galactic latitudes ($b<-0.2^\circ$ or $b>0.1^\circ$) are excluded for high-lighting the CMZ region.}
\label{fig:gdnes}
\end{center}
\end{figure*}

\begin{figure}
\begin{center}
\includegraphics[scale=0.5]{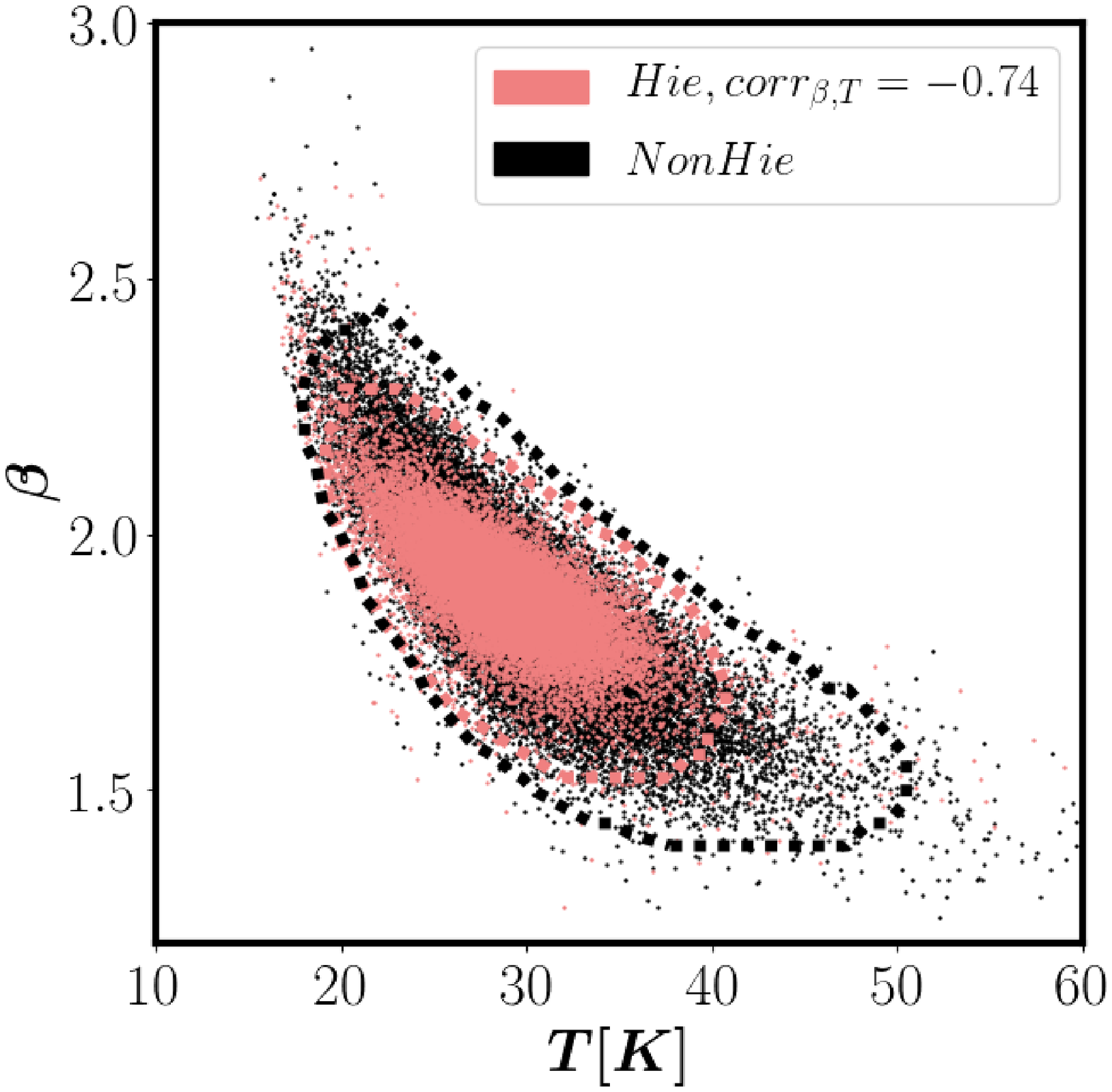}
\caption{$T-\beta$ distribution derived from a hierarchical Bayesian analysis (coral pink), compared with that derived from a non-hierarchical Bayesian analysis (same as Figure~\ref{fig:corr}, black). Both analyses are based on the STMB model with no fore/background subtraction. The contours are at the $3\sigma$ confidence level. By modeling the intrinsic $ln(T)-\beta$ distribution hierarchically as a multivariate Student-t distribution, we derive an anti-correlation: $\rho_{ln(T), \beta}=-0.74$.}
\label{fig:hie_singT}
\end{center}
\end{figure}

\begin{figure}
\begin{center}
\includegraphics[scale=0.5]{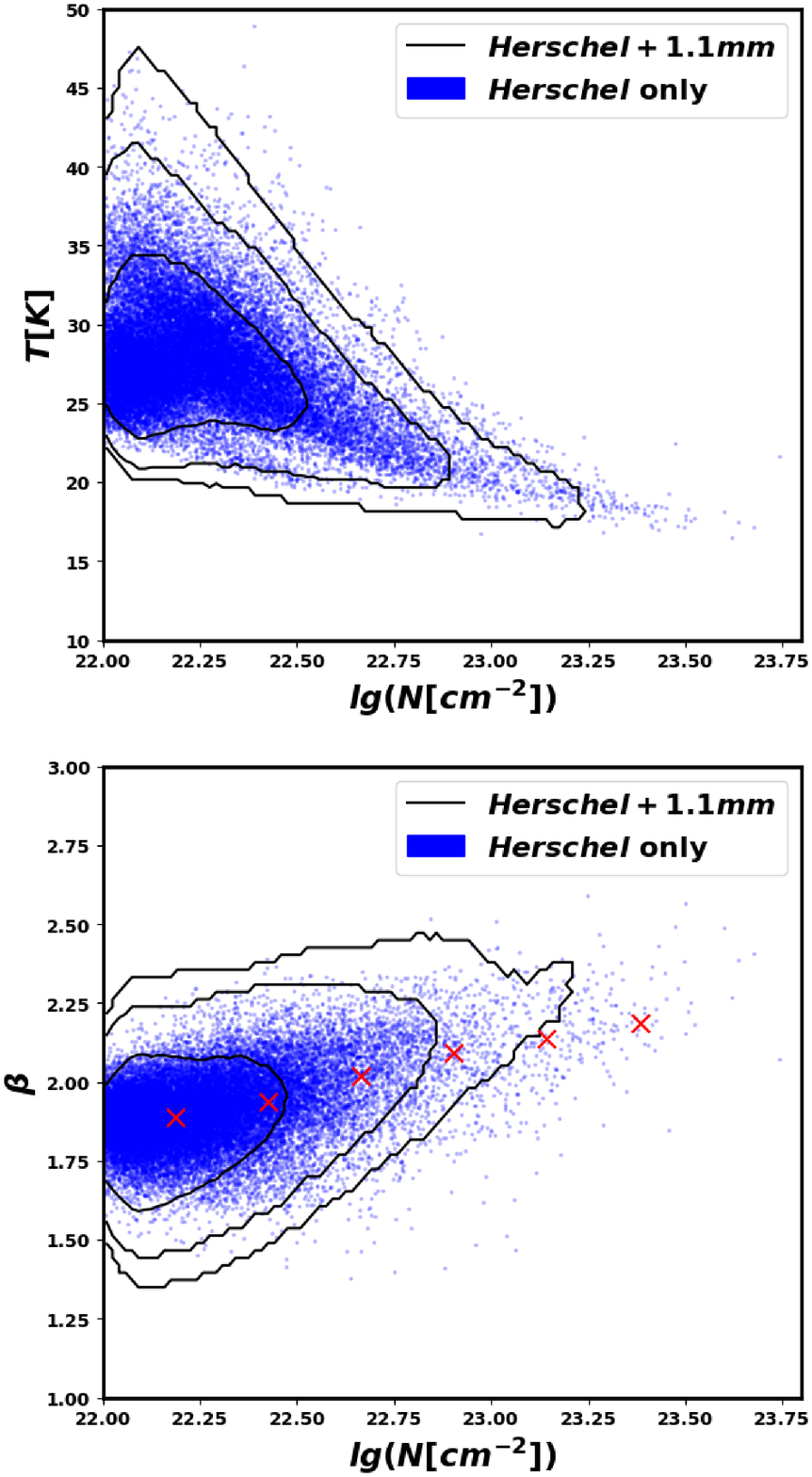}
\caption{Comparison between $N_{H_2}$, $T$ and $\beta$ derived from the \textit{Herschel}+1.1 mm maps (3-level contours from 1-3 $\sigma$) and those derived from the \textit{Herschel} $160-500$ $\mu m$ maps only (blue dots). In the lower panel, the red crosses show median $\beta$ values of the \textit{Herschel}-only results. While both results show similar trends of increasing $\beta$ toward density peaks, $\beta$ derived from the \textit{Herschel} $160-500$ $\mu m$ maps only have systemically lower values.}
\label{fig:maps_1100remv}
\end{center}
\end{figure}

In Figure~\ref{fig:gdnes} we plot the histograms of the ratios between the best-fit flux and the observed flux, $F_{fitted}$/$F_{observed}$. There is a systematical offset of $\approx10\%$ at 1.1 mm. This large systematic offset in contrast with \textit{Herschel} bands is partially due to more substantial uncertainties at 1.1 mm. It is not clear whether this systematic offset is model-driven or due to calibration error.
We do see a small systematic difference between high-density cells and low-density cells, which is however contrary to what is anticipated if the variation of $\beta$ is related to the filtering effect in the Bolocam and the AzTEC maps at 1.1 mm. Indeed, dense clouds should be less affected by the filtering effect.
As demonstrated in Figure~\ref{fig:maps_1100remv}, with the 1.1 mm map being removed, we perform the same analysis on the \textit{Herschel} $160-500$ $\mu m$ maps and find a distribution of $\beta$ similar to but systematically smaller than that in Figure~\ref{fig:maps_singT}. 

We also notice that some identified foreground objects in the CMZ show no sign of elevated $\beta$. \cite{deguchi12} suggest that the dark cloud G359.94+0.17 is composed of two clouds in the foreground, 
with $V_{lsr}=0$ $km/s$ and $15$ $km/s$. The comet-like feature near Sgr C complex ($l=359.64$, $b=0.24$)
is associated with a foreground HII region RCW 137 \citep{russeil03, tanaka14} 1.8 kpc away.
These two regions are marked in Figure~\ref{fig:maps_singT}. 

The CMZ is moderately optically thick at $160\mu m$, partially due to high column densities in the CMZ and partially due to the steep slope of the dust absorption curve. The highest optical depth is $\tau_{160} \approx 1$. $\tau_{160}$ 
is irrelevant to our choice of the amplitude of $\kappa_0$, 
since $\kappa_0$ and $N_{H_2}$ are completely degenerate. 

\section{Discussion} \label{sec:discussion}

\subsection{Increased $\beta$ in Dense Clumps}

\begin{figure}

\centering
\includegraphics[scale=0.4]{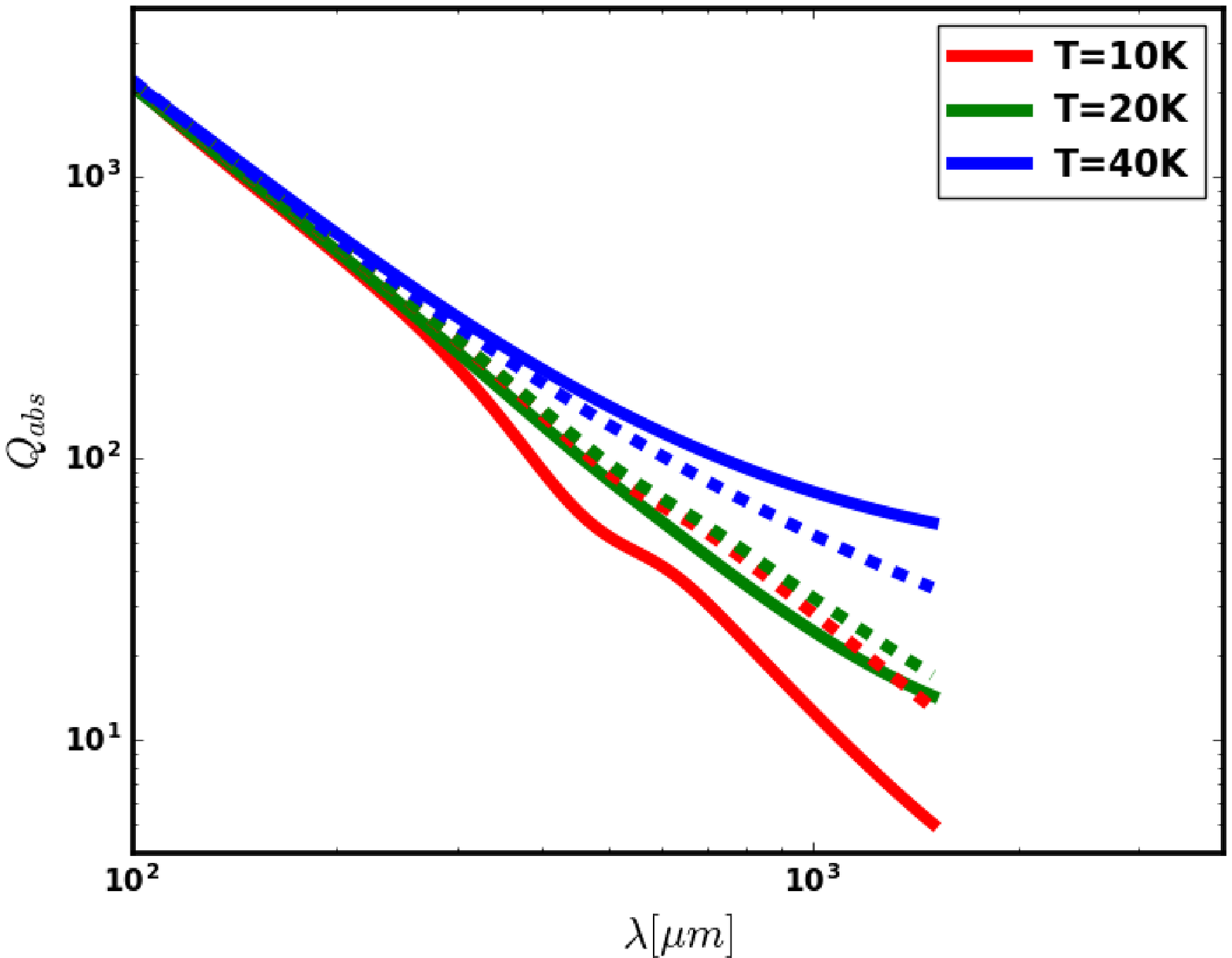}

\caption{The dust absorption curves at different temperatures, derived from a TLS model \citep{paradis14}, for the diffuse medium (dashed line) and the cold dense environment (solid lines). The power-indices $\beta$ from $500-1100$ $\mu m$ are in the range of 1.3-1.7 for the former and 1.0-2.1 for the later.}

\label{fig:Qtls}
\end{figure}

We have identified a positive-correlation between $N_{H_2}$ and $\beta$. While this trend is qualitatively in agreement with existing observations \citep{dupac03, paradis11, juvela15},  
increased $\beta$ up to 2.4 towards density peaks can not be easily
explained by existing dust models. \cite{lis98} reported such a steep absorption curve in the CMZ based on ISO observations. The origin of this trend deserves some discussion.

We noticed that in a recent study of the dust in the CMZ, \cite{arendt19} reported a null detection of any correlation between T and $\beta$, based on an analysis of the $160-500\mu m$ data from the Hi-Gal survey, same as what we use here. However, this null detection could be due to their coverage of a much larger region, extending from $l=[-1.2, 1.2]$, $b=[-0.5, 0.5]$, to both higher and lower latitudes and beyond SgrB2 in the positive Galactic longitude direction. Our study of correlations are confined to a smaller region of $l=[-0.7, 0.6]$, $b=[-0.2, 0.1]$. In fact, their results show likely a noticeable positive correlation between $N_{H_2}$ and $\beta$ (their Figure 5) in our region of interest.   

Previous studies on the molecular cloud ``Brick'' \citep{marsh16, rathborne15} adopted a fixed $\beta=1.2$ for modeling dust emission, which is significantly lower than our results. 
This low value of $\beta$ was proposed based on a comparison between the \textit{Herschel} $500$ $\mu m$ map and the Atacama Large Millimeter/submillimeter Array (ALMA) 3 mm dust continuum by \cite{rathborne14}, who find that by adopting $\beta=1.2$, the scaled \textit{Herschel} $500$ $\mu m$ map best recovers the missed large scale emission at $3$ $mm$ in the spatially filtered ALMA map. This comparison was not quantitatively detailed, and the uncertainty is not clear. Their use of lower $\beta$ should be considered an assumption instead of a measurement. 
Indeed, contrary to our results, a flattening of dust spectral index in the millimeter portion of the SED has been reported by  \textit{Herschel} and \textit{Planck} studies in some environments \citep{goldsmith97, planck11}. The origin of this flattening is not clearly understood; potential candidates are discussed in \cite{planck11}, including 1) an extra cold dust component; 2) dust growth in very dense clouds; 3) magnetic dipole emission; and 4) low energy transitions in amorphous solids. 


The total column densities we measured are similar to those derived from recent studies \citep{longmore12, rathborne15, arendt19}. The major sources of bias/uncertainty in our analysis include the assumption of the metallicity, the single temperature approximation, and the variation of the dust absorption curve. The metallicity in the Galactic Center is probably twice higher than the solar metallicity that we have assumed here \citep{shields94, najarro09}. The column densities are likely underestimated by a factor of < 2 using a single temperature approximation (Tang et al.2020, in preparation). Finally, as we'll show later in Section 5.2, the variation of the dust absorption curve could potentially lead to an overestimate of the column densities by a factor of $\approx 2$. 

It is not a trivial task to recover the intrinsic $T-\beta$ relation. \cite{shetty09} have discussed spurious correlation due to temperature mixing along the line of sight. However, this effect is more likely to suppress $\beta$ with additional cold components on the Rayleigh-Jeans tail, which cannot explain the increase of $\beta$ in the dense clouds. Our data sample the Rayleigh-Jeans tail down to $1.1$ $mm$, where the spectral slope depends only weakly on the temperature. Therefore, we conclude that the observed anti-correlation is largely intrinsic, as confirmed by our hierarchical Bayesian analysis.

\subsection{Dust Model Predicting Higher $\beta$ in Dense Regions}

The spectral index of dust absorption is expected to be environment dependent, e.g., due to  dust growth (via accretion \& coagulation, \cite{kruegel94, ossenkopf94}), or to dust destruction (e.g., shattering \& sputtering, \cite{draine79}). In dense molecular clouds, dust growth is usually expected due to high-frequency collision \& sticking with low relative velocities. Classic models of dust growth suggest that this process leads to a lowering of $\beta$ in submillimeter/millimeter wavelengths \citep{ossenkopf94, kohler12, ysard12, ysard13}. A recent model developed by \cite{jones13, kohler15} with updated optical properties of hydrogenated carbon grains could, however, reproduce the increase of $\beta$ from FIR to submillimeter ($\gtrsim 500$ $\mu m$) by introducing a new population of small hydrogenated carbon grains. This results from a transition from aromatic-rich (i.e. hydrogen poor) hydrocarbons to aliphatic-rich (hydrogen-rich) ones in dense regions. For large grains, UV photo-process can at most aromatize down to a $\approx 20$nm depth. Aliphatic-rich carbon grains have almost negligible emissivity in FIR-millimeter comparing to aromatic-rich ones and silicate grains. As a result, the spectral index of large grains is dominated by silicate features. Still, this model does not suggest $\beta$ as high as $\gtrsim 2$. It is also questionable that dust growth could occur in the dense region in the CMZ, where the turbulent velocity dispersion is enhanced by a factor of a few \citep{shetty12, kauffmann17}. Recently, \cite{hankins17} used DUSTEM to study the $3.6-70$ $\mu m$ dust SEDs of the Arched Filaments in the CMZ, and suggest a depletion of large dust grains, which is in line with our finding that there is a millimeter deficit instead of an excess.

\subsection{Dust Model Predicting a $\beta$-T Anti-Correlation}

Laboratory experiments on ``astrophysically relevant dust analogs'' suggest complex relationships between the FIR-mm spectral index and the chemical composition or the physical structure (e.g., amorphous v.s. crystalline) of dust grains \citep{boudet05, coupeaud11, demyk17A}. In these studies, an anti-correlation between T and $\beta$ for amorphous dust is commonly reported. This correlation could also be reproduced by the TLS (two-level system) model proposed by \cite{meny07}, who adopt a disordered charge distribution (DCD) on the nanometer scale and two-level systems on the atomic scale to describe the optical properties of dust. The absorption due to the DCD process is temperature independent and the combined absorption due to the TLS process, including resonant absorption, tunneling, and hopping, increases with temperature. This model is later applied by \cite{paradis11, paradis14} to successfully reproduce the SEDs of ultracompact HII regions and cold clouds observed with \textit{Herschel}/PACS \& SPIRE and CSO/Bolocam. Both \cite{paradis14} and \cite{juvela15} reported an anti-correlation between T and $\beta$ from large samples of cold clouds, which suggests that dust growth is at least not always a dominant factor in determining the spectral index. Our results confirm that this anti-correlation still exists in the more extreme CMZ environment.

In the TLS model, the unnormalized absorption coefficient $Q_{abs}$ can be divided into four components \citep{meny07, paradis11}:
\begin{equation}
Q_{abs} = Q_{DCD} + A(Q_{res}+Q_{phon}+Q_{hop}) 
\end{equation} 
\noindent where $A$ is a material-dependent parameter determining relative amplitudes of the temperature-independent DCD and the temperature-dependent TLS terms. The TLS terms, which become more important at long wavelengths, are further divided into three terms: resonant absorption $Q_{res}$, phonon-assisted tunneling relaxation $Q_{phon}$ and hopping relaxation $Q_{hop}$. A simplified TLS model provided by \cite{paradis14} reduces the dust absorption curve to a function of only temperature $T$ and wavelength $\lambda$: $Q_{abs}=Q_{abs}(\lambda, T)$, with material-dependent parameters $A$, $l_c$, and $c_{\Delta}$ determined from FIR to millimeter SEDs, separately, for two samples: a sample representing the diffuse medium (FIRAS/WMAP) and a sample representing the cold dense environment (Archeops). The charge correlation length, $l_c$, controls asymptotic behaviors of $Q_{DCD}$, which approaches $\lambda^{-2}$ at short wavelengths and approaches $\lambda^{-4}$ at long wavelengths \citep[eq 10]{paradis11}. $c_{\Delta}$ is an additional parameter of tunneling states in $Q_{hop}$ \citep[eq 14]{paradis11}. The values and uncertainties of $A$, $l_c$, and $c_{\Delta}$ in both environments are listed in Table 3 of ~\cite{paradis14}. The corresponding absorption curves for the cold dense environment are shown in Figure~\ref{fig:Qtls}. The power-indices $\beta$ from $500-1100$ $\mu m$ are in the range of 1.7-1.3 for the diffuse medium and 2.1-1.0 for the cold dense environment, decreasing with increasing temperature, from $T=10-40$ $K$.

\begin{figure*}
\begin{center}
\includegraphics[scale=0.45]{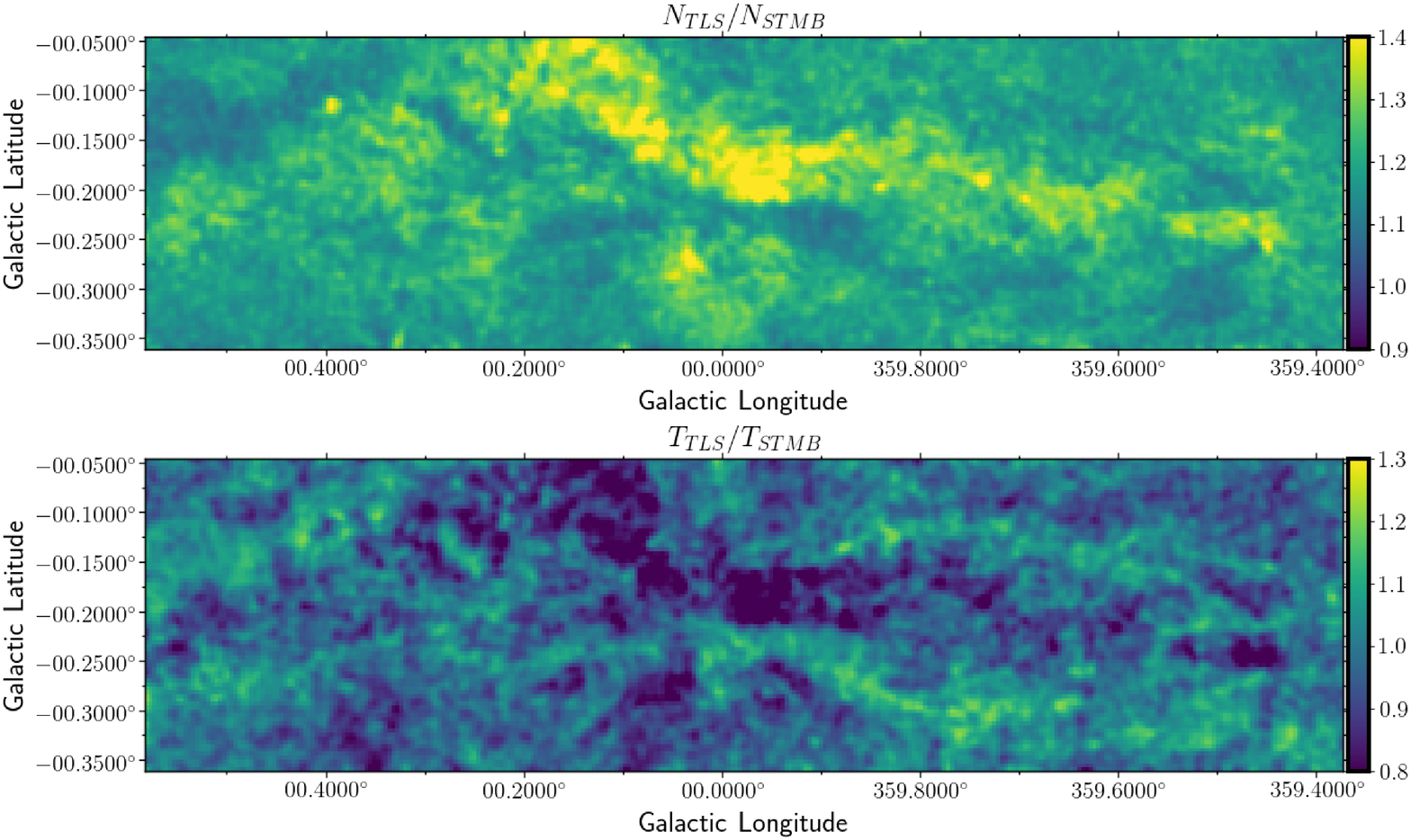}
\includegraphics[scale=0.28]{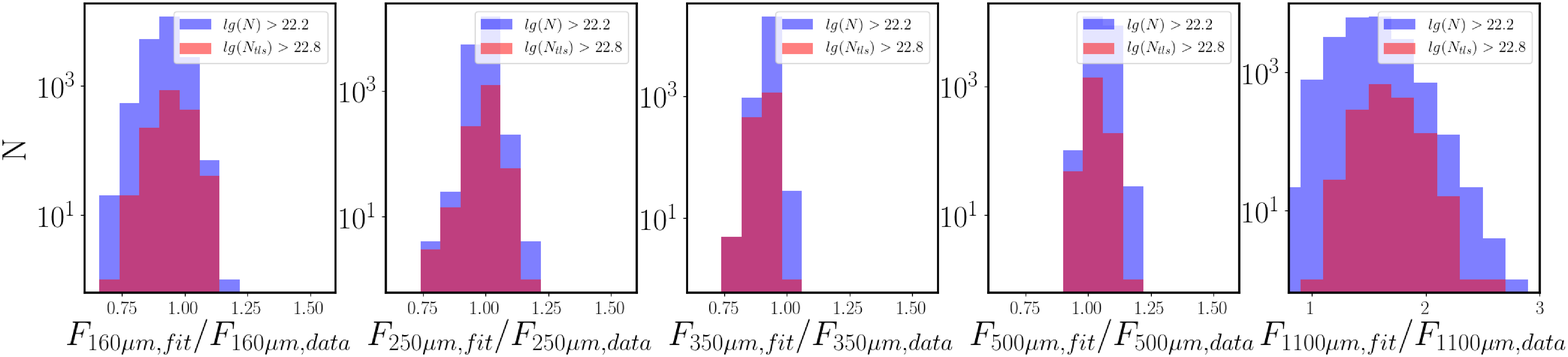}

\caption{Upper Two Panels: The column density and temperature ratios of the TLS and STMB models. Parameters related to intrinsic dust physics in the TLS model are constrained using \textit{Herschel} observations toward the Galactic cold dense environments. Lower Panels: The flux ratios $F_{fitted}$/$F_{observed}$ for best-fit TLS models. Blue bars correspond to $lg(N_{H_2}[cm^{-2}])>22.2$ pixels, while red bars to pixels with $lg(N_{H_2}[cm^{-2}]>22.8)$ only. High Galactic latitude regions ($b<-0.2^\circ$ or $b>0.1^\circ$) are not included here.}

\label{fig:map_tls}
\end{center}
\end{figure*}

We can test to what extent the TLS model could be used to describe the dust emission in the CMZ. Here we focus on the region between $-0.2^\circ<b<0.1^\circ$, where the gas density is the highest, and adopt the TLS model for grain properties in the dense environments (solid lines in Figure~\ref{fig:Qtls}). Figure~\ref{fig:map_tls} shows the best-fit maps and illustrates the goodness of the fitting. The intensity deviation of the data from the best-fit model to the observations is most significant in the 1.1 mm band, with a factor close to $50\%$ deficit. Apparently, the TLS model can not achieve higher $\beta$ and lower temperature: $\beta_{500\mu m-1.1mm}$ is at most 2.1 at a temperature of 10K and 1.8 at 20K. At shorter wavelengths, between $100 - 300$ $\mu m$, the TLS model shows a very weak dependence on temperature. These deviations, which may partly due to the temperature
mixing along the line of sight, might suggest a further difference between the grain properties in the CMZ and those in a typical Galactic dense environment. \\

The mass absorption coefficient of dust grains can be lowered at millimeter wavelengths if they have a crystalline structure~\citep{agladze96,henning97}. In crystalline material, only a small number of phonons (lattice vibrations) can contribute to FIR absorption. The disorder of the atomic arrangement in amorphous materials leads to a breakdown of the selection rules for the frequency/wavenumber that govern the excitation of vibrational modes, which induces in the longest wavelength range a broad absorption band. This difference between crystalline and amorphous materials could be observed in a broad temperature range, between $\approx 10-300$K~\citep{mennella98}. However, the formation of crystalline dust grains usually requires condensation or annealing with $T \gtrsim 1000$ $K$, and crystalline silicate, with identifiable spectral features, and are thus expected to occur primarily in the circumstellar environment, occasionally in diffuse ISM where shocks are present \citep{wright16}. Furthermore, it is expected that crystalline dust undergoes amorphization in the ISM environment \citep{kemper04}. In principle, strong shocks which prevail in the CMZ could potentially produce temperatures high enough for crystallization, but this scenario is yet to be explored.  
We conclude that the observed $T$-$\beta$ anti-correlation could be explained by four non-mutually-exclusive possibilities. 1) dust growth impediment and shatterings in the turbulent CGM environment; 2) an intrinsic $T$-$ \beta$ anti-correlation and 3) dust growth involved with hydrogenated carbons. A combination of 3) with either 1) or 2) remains the most plausible scenario.

\begin{figure*}
\begin{center}
\includegraphics[scale=0.5]{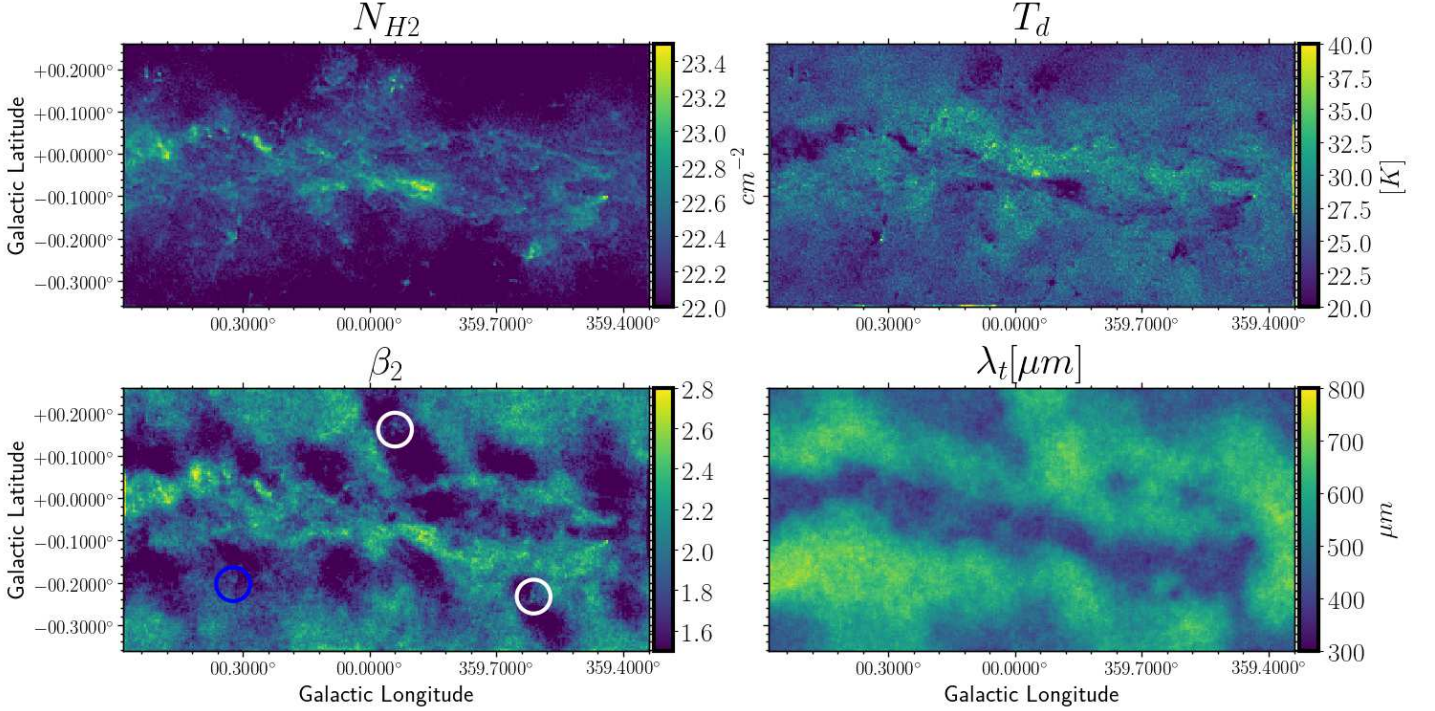}
\caption{Best-fit parameter distributions of a model with a broken power-law $\beta$ curve. 
Upper left: equivalent hydrogen column density. Upper right: dust temperature. Lower left: $\beta_2$. Lower Right: transition wavelength $\lambda_{t}$. $\beta_1$ is fixed to 2.0. The white circles again show two objects identified as foreground objects by their $V_{lsr}$, while blue circle marks an object possibly associated with the HII region SH-20.}
\label{fig:maps_bkbeta}
\end{center}
\end{figure*}

\begin{figure*}
\begin{center}
\includegraphics[scale=0.4]{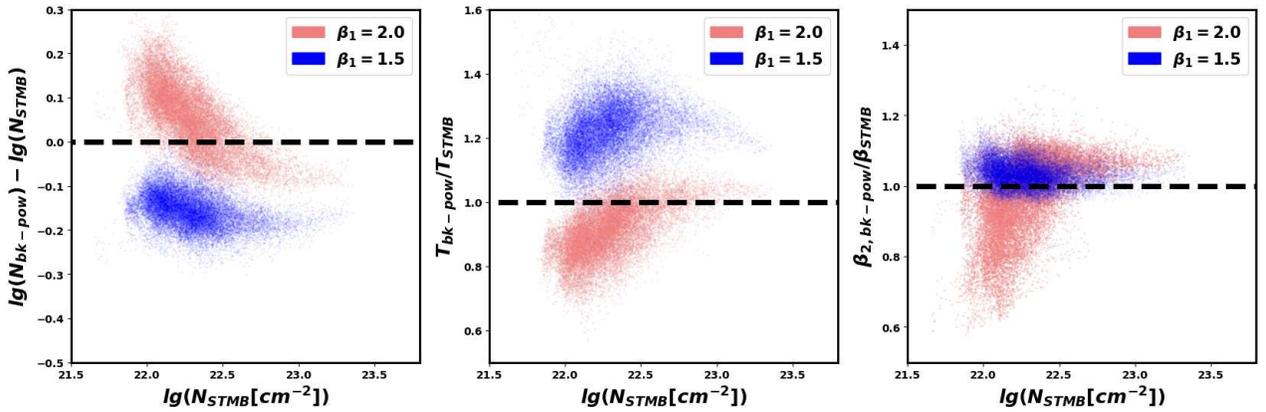}
\caption{
Comparison between the best-fit parameters of the models with the single or broken power-law $\beta$ curve: column density (left), temperature (middle) and $\beta$ (right). The label ``bk-pow'' stands for ``broken power-law''. This comparison is limited to low Galaxy latitudes: $-0.2<b<0.1$. The parameter maps are smoothed with a Gaussian kernel of $\sigma=1$ pixel before the comparison. The drop-off of $\beta_2$ with $\beta_1=2.0$ at low densities is driven by artifical stripes visible in the $\beta_2$ map in Figure~\ref{fig:maps_bkbeta}, which is casued by subtraction of CO J=3-2 in the \textit{Planck} map (see Section 3.4.1 in Paper I).  Similar drop-off is not seen in the case of $\beta_1=1.5$ because $\lambda_{t,1.5}$ is systematically smaller: $\lambda_{t,1.5}=300-400\mu m$. In comparison, $\lambda_{t,2.0}=400-700\mu m$. In the later case, $\beta_2$ is predominately determined by the 1.1 mm combined map.}
\label{fig:NT_bkbeta}
\end{center}
\end{figure*}

\subsection{Impacts of $\beta$ Variation} \label{sec:bkpw}

Both the dust growth scenarios \citep{ossenkopf94, kohler12, ysard12, ysard13, jones13, kohler15} and the TLS model \citep{meny07, paradis11, paradis14} suggest a wavelength-dependent change of $\beta$. It is, therefore, worthwhile to explore the deviations from a single power-law absorption curve. Consider a smoothly broken power law for $\kappa_{\lambda}$:

\begin{equation}\label{eq:beta_evo2}
\kappa_{\lambda} = \kappa_{\lambda_t} (\frac{\lambda}{\lambda_t})^{-\beta_1} \left\{ \frac{1}{2} \left[ 1 + (\frac{\lambda}{\lambda_t})^{\frac{1}{\delta}} \right] \right\}^{(\beta_1 -\beta_2)\delta}
\end{equation}

For simplicity, we fix $\delta$ to 0.1, this leads to a relatively sharp transition from $\beta_1$ to $\beta_2$ at wavelength $\lambda_t$. Limited by the five-bands of our SEDs, $\beta_1$ is almost completely degenerate with the dust temperature. We thus fix $\beta_1$ to a value of either 1.5 or 2.0 to investigate its impact on the measurement of the temperature and the column density. The term $\kappa_0 (\frac{\nu}{\nu_0})^{\beta}$ in Eq~\ref{eq:tau} is replaced by Eq~\ref{eq:beta_evo2}. The STMB model with this new $\beta$ curve is fitted to the CMZ maps, again we apply smoothness priors to the parameter grids to avoid over-fitting. We set $\sigma_{lg(N_{H_2})}=0.1$, $\sigma_{\beta_2}=0.2$ and $\sigma_{\lambda_t}=20$ $\mu m$. The best-fit parameter distributions for the model with $\beta_1=2.0$ are shown in Figure~\ref{fig:maps_bkbeta}. We find that $\beta_2=2-3$ and $\lambda_t \approx 500$ $\mu m$ across the map, which is within the range of those suggested by recent experimental studies on astrophysically relevant dust analogs \citep{boudet05, coupeaud11, demyk17A, demyk17B}, a summary of which is given in Table 1 of \cite{demyk13}. Figure~\ref{fig:NT_bkbeta} provides a comparison between the temperature and column densities derived from the best-fit models with the single $\beta$ and with a broken power-law $\beta$ curve. This comparison illustrates that the estimation of the temperature is very sensitive to the assumption of $\kappa_{\lambda}$. The temperatures derived with $\beta_1=1.5$ are systematically higher by $20-50\%$ compared to those derived with a $\beta_1 =2.0$, as a result of the reduced absorption coefficient at short wavelengths, which also leads to lower optical depths at $160$ $\mu m$. Still, this effect can not fully explain the discrepancy between the dust temperature and the gas temperature in the CMZ, which is a factor of 2-5 \citep{ginsburg16, krieger17}.

\section{Summary} \label{sec:conclusion}

To explore dust properties in the CMZ, we have combined the AzTEC $1.1$ $mm$ map with existing \textit{Herschel}, Plank and Bolocam surveys from $160$ $\mu m$ to $1.1$ $mm$ and carried out a joint SED analysis. We have developed an MCMC analysis tool which incorporates the knowledge of the PSFs to improve the spatial resolution, as well as the treatment of global background emission in different bands Equipped with this technique, we have explored the spatial variation of the column density, the dust temperature, and the dust spectral index $\beta$ in the CMZ. Our main results and conclusions are the following:\\

1) The spectral index $\beta$ of the dust absorption curve increases from $1.8$ to $2.4$ from intermediate column densities ($N_{H_2} \approx 22.5$ cm$^{-2}$) to high densities ($N_{H2} \approx 23.5$ cm$^{-2}$). We confirm with a hierarchical Bayesian analysis that this correlation is not due to model degeneracy. We also derive a similar distribution of $\beta$ by only using \textit{Herschel}/\textit{Planck} maps. Furthermore, we notice an absence of increased $\beta$ toward foreground dense clouds in the same field. Therefore, the increase of $\beta$ towards cold and dense clumps is induced by the CMZ environment.   \\

2) The positive correlation between $N_{H_2}$ and $\beta$ can be qualitatively, but not yet quantitatively, explained by contemporary dust models. This correlation could also be partially owing to a lack of dust growth, or even shattering due to the grain-grain collisions in a highly turbulent environment.
In principle, the correlation could be caused by an intrinsic dependence of $\beta$ on the temperature of dust. However, We find that the required dependence cannot be reproduced by either the dust growth model \citep{kohler12, kohler15} or the TLS model \citep{meny07, paradis14}. \\

3) The inferred dust temperature is strongly dependent on the assumed dust absorption curve. We show that, different assumptions for $\beta$ ($\lambda < 500\mu m$) result in $0.1-0.2$ dex difference in column and up to $50\%$ difference
in temperature. This model uncertainty is too small to be responsible for the decoupling between the gas temperature and the dust temperature observed in the CMZ (e.g., \cite{krieger17}). 

\section*{Data Availability}
The AzTEC data and products underlying this article are available at \footnote{\url{https://github.com/tangyping/products.git}}. The \textit{Herschel} Hi-Gal products were provided by the Hi-Gal team by permission. The datasets from \textit{Planck} telescope were derived from \footnote{\url{https://pla.esac.esa.int/}}. The datasets from CSO/Bolocam were derived from \footnote{\url{https://irsa.ipac.caltech.edu/data/BOLOCAM_GPS}}.

\section*{Acknowledgement}

We thank the referee for his constructive suggestions that improved this paper. The AzTEC instrument was built and operated through support from NSF grant 0504852 to the Five College Radio Astronomy Observatory. The authors gratefully acknowledge the many contributions of David Hughes in leading the LMT to its successful operational state. This work is partly supported by NASA via the grant NNX17AL67G. This work is also based on observations made with \textit{Herschel}, \textit{Planck} and the Caltech Submillimeter Observatory (CSO) telescope. \textit{Herschel} is a European Space Agency cornerstone mission with science instruments provided by European-led Principal Investigator consortia and with significant participation by NASA. \textit{Planck} is a project of the European Space Agency with instruments funded by ESA member states, and with special contributions from Denmark and NASA. CSO was operated by the California Institute of Technology under cooperative agreement with the National Science Foundatio.

\end{document}